\documentclass[lettersize,journal,twoside]{IEEEtran}
\usepackage[caption=false,font=normalsize,labelfont=sf,textfont=sf]{subfig}
\PassOptionsToPackage{ruled, linesnumbered}{algorithm2e}
\usepackage{amsmath,amsfonts,amssymb,array,graphicx,verbatim,textcomp,stfloats,url,xcolor,bm,xstring,mathrsfs,algorithm2e,cite}
\newtheorem{theorem}{Theorem}
\newtheorem{proposition}{Proposition}
\newtheorem{remark}{Remark}
\def\b#1%
{\IfSubStr{ABCDEFGHIJKLMNOPQRSTUVWXYZabcdefghijklmnopqrstuvwxyz}{#1}
	{\mathbf{#1}}
	{\bm{#1}}}
\allowdisplaybreaks      
\newcommand{\herm}{^{\mathsf{H}}}
\newcommand{\trans}{^{\mathsf{T}}}

\DeclareMathOperator{\tr}{\mathsf{Tr}}
\DeclareMathOperator{\vecd}{\mathsf{vec_d}}

\DeclareMathOperator{\diag}{\mathsf{diag}}
\DeclareMathOperator{\maximize}{\text{maximize}}
\DeclareMathOperator{\st}{\text{subject~to}}
\DeclareMathOperator{\argmin}{\mathrm{argmin}}

\begin{document}
	
	\title{\LARGE{STARS-Enabled Full-Duplex Two-Way mMIMO System Under Spatially-Correlated Channels}}
	\author{Anastasios~Papazafeiropoulos, Pandelis~Kourtessis, Symeon~Chatzinotas, Dimitra I. Kaklamani, 			Iakovos S. Venieris
		\thanks{Anastasios Papazafeiropoulos is with the Communications and Intelligent Systems Research Group, University of Hertfordshire, Hatfield AL10 9AB, U. K., and with SnT at the University of Luxembourg, Luxembourg (e-mail:tapapazaf@gmail.com). \par 
			Pandelis~Kourtessis is with the Communications and Intelligent Systems Research Group, University of Hertfordshire, Hatfield AL10 9AB, U. K. (e-mail: p.kourtessis@herts.ac.uk). \par 
			Symeon~Chatzinotas is with the SnT at the University of Luxembourg, Luxembourg (e-mail: symeon.chatzinotas@uni.lu).\par
			Dimitra I. Kaklamani is with the Microwave and Fiber Optics Laboratory, and Iakovos S. Venieris is  with the Intelligent Communications and Broadband Networks Laboratory, School of Electrical and Computer Engineering, National Technical University of Athens, Zografou, 15780 Athens,	Greece.	
	}}
	
	% The paper headers
	\markboth{Submitted to IEEE Transactions on Vehicular Technology}%
	{Papazafeiropoulos \MakeLowercase{\textit{et al.}}: STARS-Enabled Full-Duplex Two-Way \MakeLowercase{m}MIMO System Under Spatially-Correlated Channles}
	
	%\IEEEpubid{0000--0000/00\$00.00~\copyright~2021 IEEE}
	% Remember, if you use this you must call \IEEEpubidadjcol in the second
	% column for its text to clear the IEEEpubid mark.
	\renewcommand{\baselinestretch}{0.98}
	\maketitle
	
	\begin{abstract}
		Low-cost reconfigurable intelligent surfaces (RISs) \textcolor{black}{are being} considered as promising physical-layer technology for next-generation wireless networks due to their ability to re-engineer the propagation environment \textcolor{black}{by tuning their} elements. Although RIS-aided systems have many advantages, one of their major bottlenecks is that they can provide coverage only in front of the surface. Fortunately, \underline{s}imultaneous \underline{t}ransmitting \underline{a}nd \underline{r}eflecting \underline{s}urface (STARS)-assisted systems have emerged to fill this gap by providing $ 360^{\circ}$ wireless coverage. In parallel, 
		full-duplex (FD) communication offers a higher achievable rate through efficient spectrum utilization compared to the half-duplex (HD) counterpart. Moreover, two-way/bi-directional communications in an FD system can further enhance the system's spectral efficiency. Hence, in this paper, we propose a STARS-enabled massive MIMO deployment in an FD two-way communication network for highly efficient spectrum utilization, while covering the dead zones around the STARS. This model enables simultaneous information exchange between multiple nodes, while \emph{potentially} doubling the spectral efficiency (SE). By invoking the use-and-then-forget (UaTF) combining scheme, we derive a closed-form expression for an achievable SE at each user of the system considering both uplink and downlink communications based on statistical channel state information (CSI), while also accounting for imperfect CSI and correlated fading conditions. Moreover, we formulate an optimization problem to obtain an optimal passive beamforming matrix design at the STARS that maximizes the sum achievable SE. The considered problem is non-convex and we propose a provably-convergent low-complexity algorithm, termed as \underline{pro}jected \underline{gr}adient \underline{a}scent \underline{m}ethod (ProGrAM), to obtain a stationary solution. Extensive numerical results are provided to establish the performance superiority of the FD STARS-enabled system over the HD STARS-enabled and FD conventional RIS (cRIS)-enabled counterparts, and also to show the effect of different parameters of interest on the system performance.
	\end{abstract}
	
	
	\section{Introduction}
	\IEEEPARstart{R}{econfigurable} intelligent surface (RIS) has gained immense popularity in the past few years as a groundbreaking physical-layer hardware technology for beyond-fifth-generation (B5G) wireless communications~\cite{DiRenzo2020,21-TCOM-ZhangTutorial,22-Zhang-Proc}. RISs consist of a large number of sub-wavelength (nearly) passive elements made from software-controlled meta-materials~\cite{16-Nature,18-Ian-ComMag}, which makes them capable of re-engineering the wireless propagation environment. The performance superiority of RIS-aided wireless communication systems over their conventional (non-RIS-aided) counterparts has been well established in terms of spectral efficiency~\cite{22-TSP-Stefan,Pan2020,20-JSAC-maxWSR,22-Kumar-WCL}, energy efficiency~\cite{Huang2019,22-VTC-Mukherjee}, and transmit power minimization~\cite{Wu2019,22-KumarSCA-WCL} among other performance metrics.
	
	Unfortunately, in a conventional RIS (cRIS) enabled system, the elements are only able to reflect the incident electromagnetic waves (EM) to one side of the surface, leading to limited  coverage and topological constraints that restrict practical deployments. A constraint example is that the transmitter and the receiver should be located on the same side of the cRIS. This limitation has recently been overcome by a new hardware known as simultaneously transmitting and reflecting surface (STARS) which is a metasurface being capable of simultaneously transmitting and reflecting the impinging EM waves to both sides of the surface~\cite{Xu2021,Mu2021,Niu2021,Wu2021,Niu2022}. The authors in~\cite{Xu2021} provided a general hardware model and two channel models corresponding to the near- and far-field regions of STARS with two user equipments (UEs). Therein, it was shown that the coverage and diversity gain is greater than the corresponding cRIS-assisted system.  Also, in~\cite{Mu2021}, three operating protocols for adjusting the transmission and reflection coefficients of the transmitted and reflected signals from the STARS were presented, namely, energy splitting (ES), mode switching (MS), and time switching (TS). It is noteworthy that STARS has attracted significant interest not only in  academia but also in industry~\cite{DOCOMO}.

	In parallel, the in-band FD communication has gain much interest as a potential technology for B5G wireless systems due to its ability to allow both uplink (UL) and downlink (DL) operations throughout the entire frequency band which leads to a higher SE compared to its half-duplex (HD) counterpart~\cite{FDBook,22-FDMag}. Although FD is inherently impaired by loop interference (LI), this can be partially suppressed by appropriate interference mitigation techniques. It is interesting to note that although cRIS/STARS naturally supports full-duplex (FD) communications, the literature on FD cRIS-/STARS-enabled communications is rather limited. Recently, in this direction, FD has been incorporated to cRIS in~\cite{Peng2021,Shen2020,Zhang2020b,22-COMML-HardareImpaired,Deshpande2022}. More specifically, the authors in~\cite{Peng2021} obtained an optimal transmit structure and passive beamforming matrix (PBM) using a majorization-minimization (MM) algorithm to maximize the weighted sum rate for a cRIS-aided system consisting of a multi-antenna FD BS and multiple dual-antenna FD UEs. Similarly, the problem of obtaining optimal transmit structure and PBM for a cRIS-aided FD system consisting of two multi-transmit-antenna and single-receive-antenna nodes was presented in~\cite{Shen2020}. On the other hand, the problem of achievable sum rate maximization for a two-user FD cRIS-aided system where both the users were equipped with multiple transmit and receive antennas was considered in~\cite{Zhang2020b}. The analysis of the achievable sum rate in a hardware-impaired cRIS-assisted multipair FD system was presented in~\cite{22-COMML-HardareImpaired}, where the authors also proposed an optimal PBM design to maximize the sum rate using particle swarm optimization. The authors in~\cite{Deshpande2022} considered an FD multi-pair cRIS system assuming an FD BS with multiple transmit and receive antennas and multiple dual antenna FD users, while also considering the impact of correlated channels and statistical CSI availability.
	
	On the other hand, some of the recent contributions on FD STARS-enabled communications include~\cite{Papazafeiropoulos2022b,22-GLOBECOM-FD-STARS,Perera2022,22-Access-FD-STARS,22-Access-FD-STARS-TPO}. In particular, the authors in~\cite{Papazafeiropoulos2022b} considered a STARS-enabled system consisting of a two-antenna FD transmitter and two HD UEs, where they derived a closed-form expression for sum SE, and also proposed an algorithm to obtain optimal PBM design to maximize the sum SE, for both ES and MS protocols. It is noteworthy that the derived closed-form expression for SE in~\cite{Papazafeiropoulos2022b} was based on statistical channel state information (CSI) which reduces the channel estimation overhead, and the impact of channel correlation at STARS was also taken into consideration. Similarly, in~\cite{22-GLOBECOM-FD-STARS}, closed-form expressions for UL and DL outage probability were derived for a STARS-enabled system (operating in MS protocol) consisting of one FD base station (BS) equipped with two antennas, and two HD users. In~\cite{Perera2022}, the authors considered the problem of weighted sum rate maximization under quality-of-service (QoS) constraints in a STARS-enabled FD system consisting of a two-antenna FD BS and two single-antenna HD UEs, for both ES and MS protocols. The authors then reformulated the original problem to a quadratically-constrained quadratic programming (QCQP) problems using successive convex approximation (SCA) for the ES protocol and a penalty-based method for MS protocol and then obtained a solution using interior-point methods. Moreover, in~\cite{22-Access-FD-STARS}, the authors derived outage probability and ergodic rate expressions in closed-form for an FD two-way STARS-enabled device-to-device communication system, where multiple dual-antenna UEs located in the transmission regime of the STARS communicated to its paired dual-antenna UE in the reflection regime of the STARS via MS protocol. In particular, the authors partitioned the STARS elements such that only a subset of elements serve one transmitter-receiver pair, which significantly restricts the STARS capability. The problem of transmit power minimization in a STARS-enabled FD system consisting of a dual-antenna FD BS and two single-antenna HD users was presented in~\cite{22-Access-FD-STARS-TPO}, where the authors proposed an alternating optimization (AO) based algorithm to obtain optimal beamforming designs. In summary, none of the works reported in~\cite{Papazafeiropoulos2022b,22-GLOBECOM-FD-STARS,Perera2022,22-Access-FD-STARS,22-Access-FD-STARS-TPO} have considered the impact of channel correlation at the STARS (except~\cite{Papazafeiropoulos2022b}), and all are limited to an FD dual-antenna BS (except~\cite{22-Access-FD-STARS}, where the role of BS was limited to managing some control signals) and two HD users.
	
	It is noteworthy that since instantaneous CSI acquisition is a challenging task in cRIS-/STARS-aided system~\cite{22-JSTSP-Pan,22-COMST-Zhang}, consideration of instantaneous CSI in~\cite{Peng2021,Shen2020,Zhang2020b,22-COMML-HardareImpaired} for cRIS-based systems and in~\cite{22-GLOBECOM-FD-STARS,Perera2022,22-Access-FD-STARS,22-Access-FD-STARS-TPO} for STARS-based systems, where the optimization of PBM should take place at every coherence interval of the channel, is of less practical importance. On the other hand, the work presented in~\cite{Papazafeiropoulos2021,Zhi2022,Papazafeiropoulos2021b,Papazafeiropoulos2022b} rely on statistical CSI being dependent on the large-scale statistics, which change at every several coherence intervals. The design of a STARS-aided system based on statistical CSI is crucial because it significantly reduces the signaling overhead. It is worthwhile to mention that the overhead increases further in the case of the STARS~\cite{Xu2021,Mu2021}. Also, this effect becomes even more pronounced in high-speed scenarios, where the channels vary rapidly \cite{Papazafeiropoulos2023}. Nonetheless, massive multiple-input multiple-output (mMIMO) systems have been the key enabler for 5G wireless, and it is highly anticipated that mMIMO will continue to be the backbone for B5G networks~\cite{21-ComStdMag-mMIMO,21-ComMag-mMIMO}. However, none of the works have presented the performance of the STARS-aided FD two-way system with mMIMO architecture.
	
	\textit{Contributions}:  The previous observations motivate the topic of this paper. Specifically, we propose to employ STARS in an FD mMIMO  two-way network to provide full coverage for UEs in blind areas found in front and behind the STARS as depicted in Fig.~\ref{fig:SysMod}. Unlike other works in cRIS assisted two-way \textcolor{black}{mMIMO} communications, this is the only study integrating a STARS. \textcolor{black}{Note that \cite{22-Access-FD-STARS} studies a device-to-device communication system.} Moreover, we account for correlated fading and imperfect CSI. The marriage of STARS with the FD architecture improves the STARS's capability to shape the propagation environment, and thus, improves the sum SE of the system under consideration. Herein, our objective is to design the PBM at the STARS such that the sum SE of the network is maximized. The coupling of the design variables results in a non-convex optimization problem, which is challenging to solve. In this regard, we propose a projected gradient ascent method (ProGrAM), where each iteration is performed using derived closed-form expressions which maximizes the achievable \textcolor{black}{sum SE}. More specifically,
	
	\begin{itemize}
		\item  Compared to~\cite{Deshpande2022}, we have considered a STARS, while compared to \cite{Peng2021,Abdullah2020,Shen2020,Zhang2020b}, we have considered correlated fading and imperfect CSI additionally to the  STARS. \textcolor{black}{We show that in the case of spatially-correlated channels, a stronger correlation at the STARS reduces the sum SE of the network.} Also, contrary to STARS-assisted system in~\cite{Xu2021}, we have assumed multiple UEs on each side of the STARS and FD communications in a unified way.
		\item Despite that many previous works have assumed that the optimization of the PBM occurs at each coherence interval because their expressions are dependent on small-scale fading, in this paper, we optimize the PBM at every several coherence intervals since it depends on large-scale statistics. 
		\item We perform channel estimation (CE) by means of linear minimum mean-square error (LMMSE) with a low overhead compared to previous works such as~\cite{Nadeem2020} or other works that do not provide analytical expressions for the channel estimate. Also, previous works derive the individual channel estimates concerning each cRIS element and leave the inter-element correlation unknown, while we obtain analytical closed-form expressions that take into account the full STARS correlation.
		\item We derive closed-form expressions for {\color{black}the average achievable} UL and DL SE for the STARS-enabled FD two-way system with a maximal-ratio transmitter (MRT) and a maximal-ratio combining (MRC) receiver, that depend only on the path-losses and covariance matrices (large-scale statistics). To obtain the optimal PBM design, we use ProGrAM where each iteration is performed in closed form.
		\item We present extensive simulation results to establish the advantage of the STARS-aided FD system over the corresponding STARS-aided HD system and cRIS-aided FD system. \textcolor{black}{Note that the FD mode increases the difficulty in deriving the closed-form solutions compared to the HD since the former requires the derivations of the  self-interference and loop-interference terms as well as the corresponding gradients.} Moreover, we also analyze the impact of different parameters of interest on the system performance.
	\end{itemize}
	
	\paragraph*{Notations}
	Bold uppercase and lowercase letters are used to denote matrices and vectors respectively. $\b X \trans$, $\b X \herm$, $\|\b X\|$, and $\tr (\b X)$ denote the (ordinary) transpose, conjugate transpose, Frobenius norm and trace of the matrix $\b X$. A square diagonal matrix whose main diagonal consists of the elements of vector $\b x$ is represented by $\diag (\b x)$, while the elements of the main diagonal of matrix $\b X$ \textcolor{black}{are} represented by the column-vector $\vecd(\b X)$. The vectorization operator which stacks the columns of $\b X$ to create a single long column vector is denoted by $\mathsf{vec}(\b X)$. The complex-valued gradient of $f(\cdot)$ with respect to (w.r.t.) $\b X^{\ast}$ is denoted by $\nabla_{\b X} f(\cdot)$. By $\Pi_{\mathcal X}(\b x)$ we denote the Euclidean projection of $\b x$ onto the set $\mathcal X$, i.e., $\Pi_{\mathcal X}(\b x) \triangleq \argmin_{\mathbf y \in \mathcal X}\|\mathbf y - \mathbf x\|$. The vector space of all complex-valued matrices of size $M \times N$ is denoted by $\mathbb C^{M \times N}$. The amplitude and phase of a complex number $x$ are denoted by $|x|$ and $\angle x$, respectively. The statistical expectation operator is denoted using $\mathbb E \{\cdot\}$.

	%==============================
	\section{System Model}
\begin{figure}
		\centering
		\includegraphics[width=0.9\columnwidth]{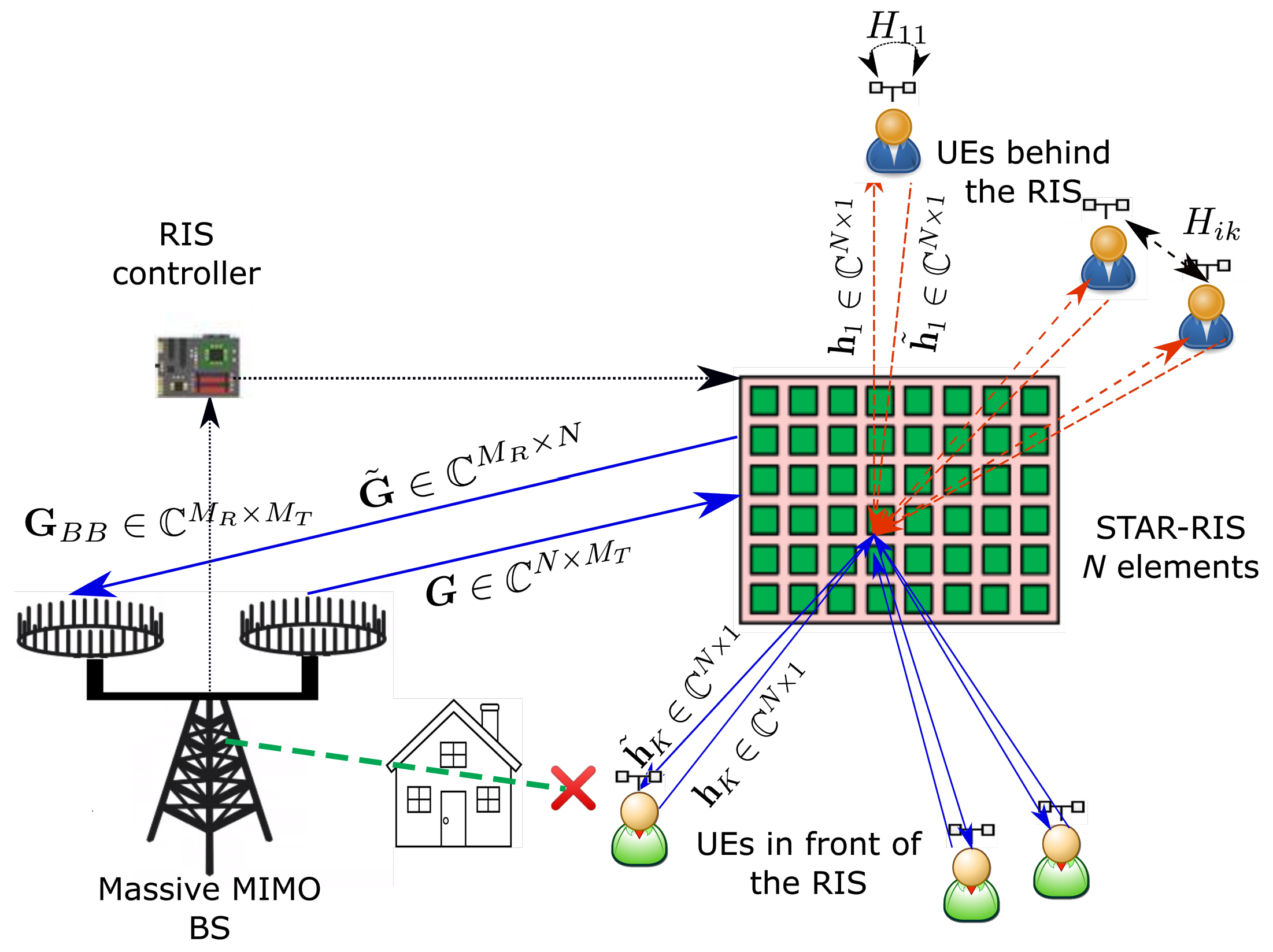}
		\caption{A STARS-Enabled FD two-way mMIMO system with multiple UEs in transmission and reflection regions.}
		\label{fig:SysMod}
	\end{figure}

	We consider a STARS-enabled FD two-way mMIMO system as shown in Fig.~\ref{fig:SysMod}, where an FD BS serves $K$ FD UEs, denoted by $\text{UE}_k, k \in \mathcal K \triangleq \{1,2,\ldots,K\}$, with the uplink and downlink transmissions occurring simultaneously at the same frequency.\footnote{\textcolor{black}{Note that this paper  considers the passive type of RIS, which is well known to suffer from the double path fading effect. To address this issue,  researchers have proposed the novel concept of active RIS \cite{Zhi2022a,Papazafeiropoulos2024}, which is equipped with amplifiers. Hence, it would be an interesting direction for future work to study active STARS with FD two-way relay.}} For this purpose, we assume that the BS is equipped with $M_{\mathrm T}$ transmit and $M_{\mathrm R}$ receive antennas, while depending on the implementation, each UE is equipped either with a single antenna or two antennas for signal transmission and reception~\cite{Sabharwal2014}. {\color{black}Without the loss of generality, we assume that the BS is located in the reflection region (i.e., in the front side) of the STARS. Moreover,} we consider the scenario where out of the $K$ UEs, $K_{\mathrm r}$ UEs are located in the reflection region and $K_{\mathrm t}$ UEs are located in the transmission region (i.e., in the back side) of the STARS, such that $K_{\mathrm r} + K_{\mathrm t} = K$. The set of indices of the UEs in the reflection and transmission regions are respectively denoted by $\mathcal K_{\mathrm r}$ and $\mathcal K_{\mathrm t}$, where $\bigcap_{\mathrm m \in \{\mathrm r, \mathrm t\}} \mathcal K_{\mathrm m} = \emptyset$ and $\bigcup_{\mathrm m \in \{\mathrm r, \mathrm t\}} \mathcal K_{\mathrm m} = \mathcal K$. Moreover, we define $\mathcal{W} = \{w_{1}, w_{2}, ..., w_{K}\} $ being the set describing the STARS operation for UEs, where $w_k = \mathrm r$ if $k \in \mathcal K_\mathrm r$ and $w_k = \mathrm t$ otherwise. Regarding the STARS, we assume that it consists of a uniform planar array (UPA) with $N \triangleq N_{\mathrm h} \times N_{\mathrm v}$ (nearly) passive elements, where $N_{\mathrm h}$ and $N_{\mathrm v}$ represent the number of columns and rows of the UPA, respectively. 
	
	Different from the half-space coverage of a cRIS, the advantage of the STARS is that it can exploit the whole space by configuring the transmitted and reflected signals in terms of two independent coefficients. In this work, we consider the ES protocol, which assumes that all STARS elements simultaneously serve all the UEs in both transmission and reflection regions. Let $s_n$ being the incident signal on the $n^{\text{th}}$ element of the STARS ($n \in \mathcal N \triangleq \{1, 2, \ldots, N\}$), the reflected and transmitted signals from the $n^{\text{th}}$ elements are respectively denoted by $\theta_{\mathrm r,n}s_n$ and $\theta_{\mathrm t,n}s_n$, respectively, where $|\theta_{\mathrm r,n}|,|\theta_{\mathrm t,n}| \in [0,1]$, $\angle \theta_{\mathrm r,n}, \angle \theta_{\mathrm t,n} \in [0,2\pi[$, and $|\theta_{\mathrm r,n}|^2+|\theta_{\mathrm t,n}|^2 = 1, \ \forall n \in \mathcal N$. Therefore, the PBM for $\text{UE}_k$ is given by $\b \Theta_{w_k} \triangleq \diag \big( \theta_{1,w_k}, \theta_{2,w_k}, \ldots, \theta_{N,w_k}\big)$.\footnote{\textcolor{black}{A more accurate model for purely passive STARS was recently proposed in~\cite{coupledSTARS}, where the transmission and reflection phase shifts are coupled with each other. However, in this paper, we consider the conventional modeling of the STARS phase shifts to keep the complexity withing tractable limits.}}
	
	Next, we describe the channel model for the system under consideration. 
	
	%-------------------------------------------------
	\subsection{Channel Model}
	Based on the narrowband quasi-static block fading channel models, we account for both small- and large-scale fading. The latter includes path-loss and STARS correlation that change slowly, i.e., the large-scale fading can be assumed to be constant for several coherence intervals contrary to small-scale fading, which remains static only during one coherence interval and then changes independently. It is worthwhile to mention that in this paper, we aim to present a unified analysis with respect to the channel estimation and data transmission phases for UL and DL scenarios. 
	
	We start by providing the channel expressions in mathematical terms. In particular, the channel matrices for the BS-STARS and STARS-BS links assuming correlation at both sides are respectively given by 
	\begin{equation} \label{GDef}
		\begin{aligned}
			\b G = & \ \sqrt{{ \delta}_{g}} \b R_{\mathrm s}^{1/2} \b D \b R_{\mathrm b}^{1/2} \in \mathbb{C}^{N \times  M_{\mathrm T}}, \\
			\tilde{\b G} = & \ \sqrt{\tilde{ \delta}_{g}}\tilde{\b R}_{\mathrm b}^{1/2}\tilde{\b D} \b R_{\mathrm s}^{1/2} \in \mathbb{C}^{M_{\mathrm R} \times  N}, 
		\end{aligned}
	\end{equation}
	where $\b R_{\mathrm{b}} \in \mathbb{C}^{M_{\mathrm T} \times M_{\mathrm T}}$, $\tilde{\b R}_{\mathrm{b}} \in \mathbb{C}^{M_{\mathrm R} \times M_{\mathrm R}}$, and $\b R_{\mathrm{s}} \in \mathbb{C}^{N \times N}$ express the \emph{deterministic} positive semi-definite correlation matrices at the BS transmit array, BS receive array, and the STARS, respectively, which are assumed to be known by the network \textcolor{black}{since they can be obtained by existing estimation methods \cite{Neumann2018,Upadhya2018}}.\footnote{\textcolor{black}{Another way of practical calculation of the covariance matrices follows. In particular,  the covariance matrices depend on the distances between the RIS elements and between the BS antennas, respectively. Moreover,  they depend on  the angles. Generally, the distances are known from the construction of the  RIS and the BS, while  the angles can be calculated when the 	locations are given.  Thus, the covariance matrices can be considered to be known.}} Note that many previous works have relied only on line-of-sight (LoS) components between the BS and the STARS, while our modeling is more general because we assume that all links are correlated Rayleigh fading distributed. Also, $\delta_{\mathrm g} $ and $\tilde{\delta}_{\mathrm g}$ represent the path-loss, and $\mathsf{vec}(\b D)\sim \mathcal{CN}\left(\b 0,\b I_{N M_{\mathrm T}}\right)$ and $\mathsf{vec}(\tilde{\b D})\sim \mathcal{CN}\left(\b 0,\b I_{ M_{\mathrm R} N}\right)$ express the fast-fading vectors. Similarly the STARS-$\text{UE}_k$ and $\text{UE}_k$-STARS channel vectors are respectively given by 
	\begin{equation} \label{hDef}
		\begin{aligned}
			\b h_{k} = & \ \sqrt{ \delta_{h_k}}  \b c_k \b R_{\mathrm{s}}^{1/2} \in \mathbb{C}^{1\times N}, \\
			\tilde{\b h}_k = & \ \sqrt{ \tilde{\delta}_{h_k}} \b R_{\mathrm{s}}^{1/2}\tilde{\b c}_k \in \mathbb{C}^{N\times 1}, 
		\end{aligned}
	\end{equation}
	where $\delta_{h_{k}}$, and $\tilde{\delta}_{h_{k}}$ represent the path-loss, and $ \b c_{k} \sim \mathcal{CN}\left(\b 0,\b I_{N}\right) $ and $\tilde{\b c}_{k} \sim \mathcal{CN}\left(\b0,\b I_{N}\right) $ express the fast-fading vectors.\footnote{\textcolor{black}{ The  analysis with correlated Rician fading that  includes a LoS component could be the topic of future work.}} Furthermore, we assume that no direct links exist between the BS and UEs due to blockages or severe path-loss. The direct link between $\text{UE}_k$ and $\text{UE}_j$ is given by $H_{kj} \sim \mathcal{CN}(0,\sigma^2_{kj}) \ \forall k,j \in \mathcal K$, and the LI channel between the transmit and receive antenna arrays at the BS is modeled as $\b G_{\mathrm b} = \tilde{\b R}_{\mathrm b}^{1/2} \bar{\b G}_{\mathrm b} \b R_{\mathrm b}^{1/2}$ with the entries in $\bar{\b G}_{\mathrm b}$ being independent and identically distributed (i.i.d.) following $\mathcal{CN}(0,\sigma^2_{\mathrm{L}})$. 
	
	As the BS is assumed to be in the reflection region, the signal traveling from the STARS to the BS is controlled by $\b \Theta_{\mathrm{r}}$. As a result, the UL cascaded channel vector for the $\text{UE}_k$-STARS-BS link $\tilde{\b u}_{k} \triangleq \tilde{\b G}\b \Theta_{\mathrm{r}} \tilde{\b h}_{k} \in \mathbb C^{M_{\mathrm R} \times 1}$ has a covariance matrix  ${\tilde{\b R}}_{k}=\mathbb E \{\tilde{\b u}_{k}\tilde{\b u}_{k}\herm\}$ given by
	\begin{align}
		\tilde{\b R}_{k}=\tilde{\delta}_{g_{k}}\tr({\b A}_{\mathrm{r}} \b \Theta_{\mathrm{r}}\herm)\tilde{\b R}_{\mathrm{b}},\label{cov1}
	\end{align}
	where $\tilde{\delta}_{g_{k}} \triangleq \tilde{\delta}_{g}\tilde{\delta}_{h_{k}} $, $ {\b A}_{\mathrm{r}} \triangleq \b R_{\mathrm{s}} \b \Theta_{\mathrm{r}} \b R_{\mathrm{s}}$, and we have used the independence between $\tilde{\b G} $ and $\tilde{\b h}_k$. Moreover, $ \mathbb E\{	\tilde{\b h}_{k}	\tilde{\b h}_{k}\herm \} = \tilde{\delta}_{h_{k}}\b R_{\mathrm{s}}$ and we have used the property $\mathbb E\{\b V \b B\b V \herm\} =\tr (\b B) \b I_{M}$ with $\b B $ being a deterministic square matrix, and $\b V$ being any matrix with i.i.d. entries of zero mean and unit variance. Similarly, the DL cascaded channel vector for the BS-STARS-$\text{UE}_k$ link ${\b u}_{k} \triangleq {\b h}_{k}\b \Theta_{w_{k}} {\b G} \in \mathbb C^{1 \times M_{\mathrm T}}$ has a covariance matrix ${{\b R}}_{k} = \mathbb E\{{\b u}_{k}\herm {\b u}_{k}\}$ given by
	\begin{align}
		{\b R}_{k}={\delta}_{g_{k}}\tr({\b A}_{w_k} \b \Theta_{w_{k}}\herm){\b R}_{\mathrm{b}},\label{cov2}
	\end{align}
	where  $\delta_{g_k} \triangleq \delta_g \delta_{h_k}$ and ${\b A}_{w_k} \triangleq \b R_{\mathrm{s}} \b \Theta_{w_k} \b R_{\mathrm{s}}$.
	It is noteworthy that in the case of no correlation at the STARS, we have $\b R_{\mathrm{s}} =\b I_{N} $, and thus $\tilde{\b R}_{k}$ and ${\b R}_{k} $ are independent of the phase shifts but depend only on the amplitudes, which was also observed in~\cite{Papazafeiropoulos2022}.
	
	In the next section, we discuss the signal transmission and reception for the uplink and downlink scenarios. 
	
	%==============================
	\section{Signal Transmission and Reception}
	During the data transmission phase, the STARS transmits and reflects the signals from the simultaneously transmitting antennas of the BS and UEs. The signal transmitted by the FD BS is given by $\b x_{\mathrm b} = \sqrt{\beta} \sum_{k \in \mathcal K} \sqrt{p_{\mathrm bk}} \b f_k s_{\mathrm bk}$, where $p_{\mathrm bk}$ is the power allocated for $\text{UE}_k$, $s_{\mathrm bk}$ is the symbol intended for $\text{UE}_k$, and $\b f_k$ is the corresponding precoding vector. Define $\b F \triangleq [\b f_1, \b f_2, \ldots, \b f_K] \in \mathbb C^{M_{\mathrm T} \times K}$ as the precoding matrix, and the scaling factor $\beta$ ensures $\mathbb E\{\b x_{\mathrm b} \herm \b x_{\mathrm b} \} = p_{\mathrm b}$, where $p_{\mathrm b}$ is the total transmit power budget at the BS. For the sake of tractability and following the mMIMO literature, we consider the equal-power allocation among the UEs, i.e., $p_{\mathrm bk} = p_{\mathrm b}/K, \ \forall k \in \mathcal K$, resulting in $\beta = K/\mathbb E\{\tr \big(\b F \b F \herm \big)\}$. On the other hand, the signal transmitted from $\text{UE}_k$ is given by $x_{\mathrm uk} = \sqrt{p_{\mathrm uk}} s_{\mathrm uk}$, where $p_{\mathrm uk}$ is the transmit power and $s_{\mathrm uk}$ is the transmitted symbol from $\text{UE}_k$. 
	With this background, the signal received at the BS is given by 
	\begin{align}
		\b y_{\mathrm b} = \tilde{\b G} \b \Theta_{\mathrm r} \b y_{\mathrm s} + \b G_{\mathrm b} \b x_{\mathrm b} +\b z_{\mathrm b}, \label{BS_received}
	\end{align}
	where $\b y_{\mathrm s}$ is the signal received at the STARS due to the signals transmitted from the BS and UEs, the second term accounts for the BS LI, and $\b z_{\mathrm b} \sim \mathcal{CN}(\b 0, \sigma^2_{\mathrm b} \b I_{M_{\mathrm R}})$ is the additive White Gaussian noise (AWGN) at the BS. The signal received at the STARS is given by 
	\begin{equation}
		\b y_{\mathrm s} = \b G \b x_{\mathrm b} + \sum \nolimits_{k \in \mathcal K} \tilde{\b h}_k x_{\mathrm uk}. \label{RIS_received}
	\end{equation}
	The first and second terms in the preceding equation correspond to the signal received from the BS and $K$ UEs, respectively.
	
	In this paper, we adopt linear precoding for DL and linear combining for UL communications. Specifically, for the detection of signal from $\text{UE}_k$ to BS, we use the combining vector $\b v_k$ at the BS, which yields 
	\begin{align}
		& y_{\mathrm{u}k} =\mathbf{v}_{k}\herm\mathbf{y}_{\mathrm{b}}\nonumber \\
		= \ & \mathbf{v}_{k}\herm\bigg[\tilde{\mathbf{G}}\boldsymbol{\Theta}_{\mathrm{r}}\mathbf{G}\sqrt{\dfrac{\beta p_{\mathrm{b}}}{K}}\sum\nolimits _{j\in\mathcal{K}}\mathbf{f}_{j}s_{\mathrm{b}j}+\tilde{\mathbf{G}}\boldsymbol{\Theta}_{\mathrm{r}} \notag \\
		& \quad \times \sum\nolimits _{j\in\mathcal{K}}\tilde{\mathbf{h}}_{j}\sqrt{p_{\mathrm{u}j}}s_{\mathrm{u}j} +\mathbf{G}_{\mathrm{b}}\sqrt{\dfrac{\beta p_{\mathrm{b}}}{K}}\sum\nolimits _{j\in\mathcal{K}}\mathbf{f}_{j}s_{\mathrm{b}j}+\mathbf{z}_{\mathrm{b}}\bigg]\nonumber \\
		= \ & \sqrt{p_{\mathrm{u}k}}\mathbf{v}_{k}\herm\tilde{\mathbf{u}}_{k}s_{\mathrm{u}k}+\sum \nolimits_{i\in\mathcal{K}\setminus\{k\}}\sqrt{p_{\mathrm{u}i}}\mathbf{v}_{k}\herm\tilde{\mathbf{u}}_{i}s_{\mathrm{u}i}+\sqrt{\dfrac{\beta p_{\mathrm{b}}}{K}}\mathbf{v}_{k}\herm\tilde{\mathbf{G}} \nonumber \\
		& \times \boldsymbol{\Theta}_{\mathrm{r}}\mathbf{G} \sum_{j\in\mathcal{K}}\mathbf{f}_{j}s_{\mathrm{b}j}+\sqrt{\dfrac{\beta p_{\mathrm{b}}}{K}}\mathbf{v}_{k}\herm\mathbf{G}_{\mathrm{b}}\sum\nolimits _{j\in\mathcal{K}}\mathbf{f}_{j}s_{\mathrm{b}j}+\mathbf{v}_{k}\herm\mathbf{z}_{\mathrm{b}}\nonumber \\
		= \ & \underbrace{\sqrt{p_{\mathrm{u}k}}\mathbb{E}\big\{\mathbf{v}_{k}\herm\tilde{\mathbf{u}}_{k}\big\} s_{\mathrm{u}k}}_{\text{desired signal over known channel}}+\underbrace{\sqrt{p_{\mathrm{u}k}}\big(\mathbf{v}_{k}\herm\tilde{\mathbf{u}}_{k}-\mathbb{E}\big\{\mathbf{v}_{k}\herm\tilde{\mathbf{u}}_{k}\big\}\big)s_{\mathrm{u}k}}_{\text{desired signal over unknown channel}} \notag \\
		& +\sum_{i\in\mathcal{K}\setminus\{k\}}\underbrace{\sqrt{p_{\mathrm{u}i}}\mathbf{v}_{k}\herm\tilde{\mathbf{u}}_{i}s_{\mathrm{u}i}}_{\text{UL MUI}} +\underbrace{\sqrt{\dfrac{\beta p_{\mathrm{b}}}{K}}\mathbf{v}_{k}\herm\tilde{\mathbf{G}}\boldsymbol{\Theta}_{\mathrm{r}}\mathbf{G}\sum\nolimits _{j\in\mathcal{K}}\mathbf{f}_{j}s_{\mathrm{b}j}}_{\text{BS SI}} \nonumber \\
		&+\underbrace{\sqrt{\dfrac{\beta p_{\mathrm{b}}}{K}}\mathbf{v}_{k}\herm\mathbf{G}_{\mathrm{b}}\sum\nolimits _{j\in\mathcal{K}}\mathbf{f}_{j}s_{\mathrm{b}j}}_{\text{BS LI}}+\underbrace{\mathbf{v}_{k}\herm\mathbf{z}_{\mathrm{b}}}_{\text{post-combining AWGN}}, \label{rx-UL}
	\end{align}	 
	where the third term is the UL multiuser interference (MUI) from other UEs, the fourth term is the BS self-interference (SI) from its own transmit signal reflected by the STARS, and the fifth term is the BS LI. 
	
	On the other hand, the DL signal received by $\text{UE}_k$ is given by 
	\begin{align}
		& y_{\mathrm{d}k}= \mathbf{h}_{k}\boldsymbol{\Theta}_{w_{k}}\mathbf{y}_{\mathrm{s}}+\sum_{j\in\mathcal{K}}H_{kj}\sqrt{p_{\mathrm{u}j}}s_{\mathrm{u}j}+z_{\mathrm{d}k} \notag \\
		= & \ \sqrt{\dfrac{\beta p_{\mathrm{b}}}{K}}\mathbf{u}_{k}\mathbf{f}_{k}s_{\mathrm{b}k}+\sqrt{\dfrac{\beta p_{\mathrm{b}}}{K}}\sum_{i\in\mathcal{K}\setminus\{k\}}\mathbf{u}_{k}\mathbf{f}_{i}s_{\mathrm{b}i} +\mathbf{h}_{k}\boldsymbol{\Theta}_{w_{k}}\notag \\
		&\times \sum_{j\in\mathcal{K}}\sqrt{p_{\mathrm{u}j}}\tilde{\mathbf{h}}_{j}s_{\mathrm{u}j} \! + \!\!\!\!\! \sum_{i\in\mathcal{K}\setminus\{k\}} \!\!\!\!\!\! H_{ki}\sqrt{p_{\mathrm{u}i}}s_{\mathrm{u}i}+H_{kk}\sqrt{p_{\mathrm{u}k}}s_{\mathrm{u}k}+z_{\mathrm{d}k} \notag \\
		= & \ \underbrace{\sqrt{\dfrac{\beta p_{\mathrm{b}}}{K}}\mathbb{E}\big\{\mathbf{u}_{k}\mathbf{f}_{k}\big\} s_{\mathrm{b}k}}_{\text{desired signal over known chanel}}+\underbrace{\sqrt{\dfrac{\beta p_{\mathrm{b}}}{K}}\big(\mathbf{u}_{k}\mathbf{f}_{k}-\mathbb{E}\big\{\mathbf{u}_{k}\mathbf{f}_{k}\big\}\big)s_{\mathrm{b}k}}_{\text{desired signal over unknown channel}} \notag \\
		%	& + \underbrace{\sqrt{\dfrac{\beta p_{\mathrm{b}}}{K}}\sum \nolimits_{i\in\mathcal{K}\setminus\{k\}}\mathbf{u}_{k}\mathbf{f}_{i}s_{\mathrm{b}i}}_{\text{DL MUI}} +\underbrace{\mathbf{h}_{k}\boldsymbol{\Theta}_{w_{k}}\sum \nolimits_{j\in\mathcal{K}}\sqrt{p_{\mathrm{u}j}}\tilde{\mathbf{h}}_{j}s_{\mathrm{u}j}}_{\text{UE CCI}} \notag \\
		& \quad+\underbrace{\sum \nolimits_{i\in\mathcal{K}\setminus\{k\}}H_{ki}\sqrt{p_{\mathrm{u}i}}s_{\mathrm{u}i}}_{\text{IUI}}+\underbrace{H_{kk}\sqrt{p_{\mathrm{u}k}}s_{\mathrm{u}k}}_{\text{UE-LI}}+z_{\mathrm{d}k}, \label{rx-DL}
	\end{align}
	where the third, fourth, fifth, and sixth terms refer to DL MUI, UE co-channel interference, interuser interference, and $\text{UE}_k$'s LI. Moreover, $z_{\mathrm dk} \sim \mathcal{CN}(0,\sigma^2_{\mathrm dk})$ is the AWGN at $\text{UE}_k$. For ease of exposition, in the rest of this paper, we assume $\sigma^2_{\mathrm dk} = \sigma^2_{\mathrm d} \ \forall k \in \mathcal K$.
	
	Before proceeding further, we present the channel estimation for the system under consideration first, in the next section. 
	%==============================
	
	\section{Channel Estimation}
	In practice, perfect CSI is not available at the BS. Additionally, a STARS is implemented by means of nearly passive elements that cannot send pilots to the BS and allow channel estimation or receive pilots and estimate the corresponding channel. 
	
	In this direction, given that the FD BS is deployed with separate arrays for transmission and reception, it will estimate the cascaded UL and DL channels separately, in order to design the decoding and precoding vectors $ \b v_{k} $ and $ \b f_{k} $, respectively. Note that the physical separation of transmit and receive antennas is a common passive cancellation technique to combat LI, and the corresponding UL and DL channels, i.e., $ \tilde{\b G} $ and $ \b G $ are not identical~\cite{Zhang2016,Shojaeifard2017}. In parallel, each FD UE is equipped with separate transmit and receive antennas. Hence, the  DL channel between the STARS and $\text{UE}_k$ receive antenna, and the UL channel between $\text{UE}_k$ and the STARS are given by $ \b h_{k} $ and $ \tilde{\b h}_{k} $, respectively. This means that the effective DL and UL channels from the BS to the UE receive antenna and from the transmit UE antenna to the BS are not the same. Different from~\cite{Deshpande2022} and~\cite{Nadeem2020}, which assumed $N$ phases for channel estimation, we propose a scheme where channel estimation is to be performed in a single phase that achieves a higher SE due to a higher pre-log factor and can be applied not only for low-speed UEs but also for fast UEs  since the training does not take a long time. 
	\begin{figure}
		\centering 
		\includegraphics[width=0.9\columnwidth]{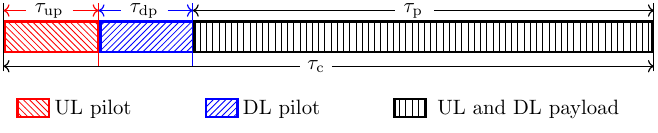}
		\caption{A coherence interval.}
		\label{Coherence}
	\end{figure}
	
	We assume that  each block has a duration of $\tau_{\mathrm{c}}=B_{\mathrm{c}}T_{\mathrm{c}}$ channel uses, where  $ B_{\mathrm{c}} $ is the coherence bandwidth, and $ T_{\mathrm{c}} $ is the coherence time. From this duration, \textcolor{black}{$\tau_{\mathrm{up}}\in \mathbb{Z}$ and $\tau_{\mathrm{dp}}\in \mathbb{Z}$ channel uses} are considered for UL and DL pilots, respectively, and $\tau_{\mathrm c} - (\tau_{\mathrm{up}}+\tau_{\mathrm{dp}})$ channel uses are considered for simultaneous transmission \textcolor{black}{for} UL and DL data, as shown in Fig.~\ref{Coherence}.
	
	During the UL training phase of duration $\tau_{\mathrm{up}}$ symbols, UEs transmit orthogonal pilot sequences to the BS via STARS. Especially, $\tilde{\b x}_{\mathrm pk}=[\tilde{x}_{\mathrm pk,1}, \tilde{x}_{\mathrm pk,2}, \ldots, \tilde{x}_{\mathrm pk,\tau_{\mathrm{up}}
	}]\trans \in \mathbb{C}^{\tau_{\mathrm{up}}\times 1}$ denotes the pilot sequence of $\text{UE}_k$ with $\tilde{\b x}_{\mathrm pk}\herm \tilde{\b x}_{\mathrm pl} = 0~\forall k,l \in \mathcal K, k\neq l$ and $\tilde{\b x}_{\mathrm pk}\herm \tilde{\b x}_{\mathrm pk} = \tau_{\mathrm{\mathrm{up}}} p_{\mathrm{Train}}$ joules with $p_{\mathrm{Train}} = |\tilde{x}_{\mathrm pk,i}|^{2} ,~\forall k \in \mathcal K,i \in \{1, 2, \ldots, \tau_{\mathrm{up}}\}$ being the common average transmit power per UE during the UL training phase.
	
	The received signal at the BS during the UL training period is given by
	\begin{align}
		\b Y_{\mathrm{u}} = \sum \nolimits_{i \in \mathcal K} \tilde{\b u}_{i}\tilde{\b x}_{\mathrm{p}i}\herm +	\tilde{\b W},\label{train1}
	\end{align}
	where $ \tilde{\b W} \in \mathbb{C}^{M_{\mathrm R} \times \tau_{\mathrm{up}}}$ is the AWGN matrix at the BS with independent columns, each one distributed as $ \mathcal{CN}\left(\b 0,\sigma_{\mathrm{u}}^2\b I_{M_{\mathrm R}}\right)$. The received training signal at the BS is multiplied with the transmitted training sequence from $\text{UE}_k$, given by~\eqref{train1}, to eliminate the interference caused by other UEs which yields
	\begin{align}
		\tilde{\b r}_{k}=\tilde{\b u}_{k} +\frac{\tilde{\b w}}{ \tau_{\mathrm{up}} p_{\mathrm{Train}}},\label{train2}
	\end{align}
	where  $\tilde{\b w}=\textcolor{black}{\tilde{\b W} {\tilde{\b x}}_{\mathrm{p}k}\sim \mathcal{CN}(\b0,\tau_{\mathrm{up}} p_{\mathrm{Train}} \sigma_{\mathrm{u}}^2\b I_{M_{\mathrm R}})} $. 
	
	\begin{proposition}\label{PropositionuplinkChannel}
		The MMSE estimate of the UL cascaded channel $\tilde{\b u}_{k}$ is given by
		\begin{align}
			\hat{\tilde{\b u}}_{k}= \tilde{\b R}_k \tilde{\b Q}_{k} \tilde{\b r}_{k},\label{estim1}
		\end{align}
		where 
		\begin{equation} 
			\tilde{\b Q}_{k} = \left( \tilde{\b R}_{k} + \frac{{\sigma}_{\mathrm{u}}^2}{  \tau_{\mathrm{up}} p_{\mathrm{Train}} } \b I_{M_{\mathrm R}} \right)^{-1},\label{QKtilde}
		\end{equation} 
		and $ \tilde{\b r}_{k}$, given by~\eqref{train2}, is the noisy observation of $\tilde{\b u}_k$.
	\end{proposition}
	\begin{IEEEproof}
		The proof follows the common steps of MMSE channel estimation~\cite{massivemimobook}.
	\end{IEEEproof}
	Based on the property of orthogonality of MMSE estimation, the perfect cascaded UL channel is given by 
	\begin{align}
		\tilde{\b u}_{k}={\hat{\tilde{\b u}}}_{k}+{\tilde{\b e}}_{k} ,\label{perfetULchannel} 
	\end{align}
	where $\hat{\tilde{\b u}}_{k}$ and ${\tilde{\b e}}_{k} $, being the estimated channel and the channel error vector, having zero mean and variances $\tilde{\b \Psi}_{k}=\tilde{\b R}_{k}\tilde{\b Q}_{k}\tilde{\b R}_{k}$ and $ \tilde{\b E}_{k}=\tilde{\b R}_{k}-\tilde{\b \Psi}_{k}$, respectively.
	
	\textcolor{black}{In a similar way, the following proposition provides the  MMSE estimate of the DL cascaded channel $ \b u_{k} $, where $w_k\sim \mathcal{CN}(0,\sigma_{\mathrm dk}^2)$ is the AWGN at $\text{UE}_k$.}
	%estimation of $ \b u_{k} $ is based on the following equation
	%\begin{align}
	%	{ r}_{k}={ u}_{k} +\frac{ w_{k}}{ \tau_{\mathrm{dp}} p_{\mathrm{Train}}},\label{rDef}
	%\end{align}
	%where {\color{black}$\b w_{k} = w_k \b x_{\mathrm pk}$ }, $w_k\sim \mathcal{CN}(0,\sigma_{\mathrm dk}^2)$ is the AWGN at $\text{UE}_k$, and $\b x_{\mathrm pk} \in \mathbb{C}^{\tau_{\mathrm{dp}}\times 1}$ is the pilot sequence associated with $\text{UE}_k$. The following proposition presents the estimated DL cascaded channel $ \hat{\b u}_{k} $.
	
	\begin{proposition}\label{PropositiondownlinkChannel}
		The MMSE estimate of the DL cascaded channel $ \b u_{k} $ is given by
		%	\begin{align}
			%		\hat{{\b u}}_{k}={\b r}_{k} {\b Q}_{k}{\b R}_{k} ,\label{estim2}
			%	\end{align}
		%	where $ {\b Q}_{k} = \left({\b R}_{k} + \frac{{\sigma}_{\mathrm{d}k}^2}{  \tau_{\mathrm{dp}} p_{\mathrm{Train}} }\b I_{M_{\mathrm T}}\right)^{-1}$, and $ {\b r}_{k}$ is given in~\eqref{rDef}. Accordingly, the perfect cascaded DL channel is given by
		\begin{align}
			{\b u}_{k}={\hat{{\b u}}}_{k}+{{\b e}}_{k} ,\label{perfectDLchannel} 
		\end{align}
		where $\hat{{\b u}}_{k}$ and ${{\b e}}_{k}$ have zero mean and variances $	{\b \Psi}_{k} = {\b R}_{k}{\b Q}_{k}{\b R}_{k}$ and $ {\b E}_{k}={\b R}_{k}-{\b \Psi}_{k}$ with $ {\b Q}_{k} = \left({\b R}_{k} + \frac{{\sigma}_{\mathrm{d}k}^2}{  \tau_{\mathrm{dp}} p_{\mathrm{Train}} }\b I_{M_{\mathrm T}}\right)^{-1}$, respectively.		
	\end{proposition}
	\begin{IEEEproof}
		The proof follows the common steps of MMSE channel estimation~\cite{massivemimobook}.
	\end{IEEEproof}
	\begin{remark}
		The proposed method has a number of advantages compared to other methods. First, it comes with lower complexity and requires less hardware complexity compared to estimating the individual channels, which is a quite demanding procedure as the number of surface elements increases. Note that if we would like to estimate the  individual channels, one solution would be to assume receive radio frequency (RF) chains integrated into the STARS as in~\cite{Zheng2021}. Second, we have achieved to derive the channel estimates in closed forms in terms of large-scale statistics. This means that channel estimation can be performed at every several coherence intervals. Other existing methods do not capture the correlation of the whole surface but they obtain the estimated channel per element~\cite{Nadeem2020}, or they do not provide analytical expressions, e.g.,~\cite{He2019}.
	\end{remark}
	
	%==============================
	\section{Problem Formulation}
	In this section, we first derive closed-form expressions for UL and DL SE, and then formulate an optimization problem to find the optimal PBM that maximizes the achievable sum SE.  \textcolor{black}{The self-interference,   loop interference, and the co-channel interference for a STAR-RIS system induces certain difficulties as can be seen by the mathematical derivations below. Notably, similar difficulties are met during the optimization of the sum SE in Sec. VI, where we obtain specific derivatives.}
	
	%---------------------------------------------
	\subsection{Average Uplink SE}
	Considering the worst-case additive noise for all the terms in~\eqref{rx-UL} except the desired signal, the {\color{black}expression for average achievable UL sum SE} using the use-and-then-forget (UaTF) technique~\cite{massivemimobook} is given by 
	\begin{equation}
		\mathsf{SE}_{\mathrm u}=\zeta\sum\nolimits _{k\in\mathcal{K}}\log_{2}\big(1+\gamma_{\mathrm{u}k}\big), \label{rate-UL-Def}
	\end{equation}
	where $\zeta=\tfrac{\tau_{\mathrm{c}}-\tau_{\mathrm{p}}}{\tau_{\mathrm{c}}}$,
	$\gamma_{\mathrm{u}k}=S_{\mathrm{u}k}/I_{\mathrm{u}k}$ with 
	\begin{equation}
		S_{\mathrm{u}k}=p_{\mathrm{u}k}\big|\mathbb{E}\big\{\mathbf{v}_{k}\herm\tilde{\mathbf{u}}_{k}\big\}\big|^{2}, \label{Suk-Def}
	\end{equation}
	and 
	\begin{align}
		& I_{\mathrm{u}k} =\mathsf{var}\big\{\sqrt{p_{\mathrm{u}k}}\mathbf{v}_{k}\herm\tilde{\mathbf{u}}_{k}\big\}+\sum \nolimits_{i\in\mathcal{K}\setminus\{k\}}p_{\mathrm{u}i}\mathbb{E}\big\{\big|\mathbf{v}_{k}\herm\tilde{\mathbf{u}}_{i}\big|^{2}\big\} \nonumber \\
		&+\dfrac{\beta p_{\mathrm{b}}}{K}\sum \nolimits_{j\in\mathcal{K}}\mathbb{E}\big\{\big|\mathbf{v}_{k}\herm\big(\tilde{\mathbf{G}}\boldsymbol{\Theta}_{\mathrm{r}}\mathbf{G}+\mathbf{G}_{\mathrm{b}}\big)\mathbf{f}_{j}\big|^{2}\big\}+\sigma_{\mathrm{b}}^{2}\mathbb{E}\big\{\big\Vert\mathbf{v}_{k}\herm\big\Vert^{2}\big\}. \label{Iuk-Def}
	\end{align}
	\begin{theorem} \label{thm:UL-SINR}
		Considering MRC decoding at the BS with $\mathbf{v}_{k}=\hat{\tilde{\mathbf{u}}}_{k}$, closed-form expressions for $S_{\mathrm uk}$ and $I_{\mathrm uk}$ are respectively given by 
		\begin{align}
			S_{\mathrm uk} = p_{\mathrm{u}k}\tr^{2}\big(\tilde{\boldsymbol{\Psi}}_{k}\big), \label{Suk-Closed}
		\end{align}
		and
		\begin{align}
			& I_{\mathrm uk} =  \tr(\tilde{\boldsymbol{\Psi}}_{k}\tilde{\mathbf{R}}_{\mathrm{sum}})+\dfrac{p_{\mathrm{b}}}{\tr(\boldsymbol{\Psi}_{\mathrm{sum}})}\tr\big(\tilde{\boldsymbol{\Psi}}_{k}\tilde{\mathbf{R}}_{\mathrm{b}}\big)\tr\big(\mathbf{R}_{\mathrm{b}}\boldsymbol{\Psi}_{\mathrm{sum}}\big) \notag \\
			& \times \!\! \Big\{\!\delta_{g}\tilde{\delta}_{g}\tr\big(\b A_{\mathrm r}\boldsymbol{\Theta}_{\mathrm{r}}\herm\big) \!+\!\sigma_{\mathrm{L}}^{2}\Big\}\!-\!p_{\mathrm{u}k}\tr\big(\tilde{\boldsymbol{\Psi}}_{k}^{2}\big)\!+\!\sigma_{\mathrm{b}}^{2}\tr\big(\tilde{\boldsymbol{\Psi}}_{k}\big), \label{Iuk-Closed}
		\end{align}
		where $\tilde{\b R}_{\mathrm{sum}} \triangleq \sum_{j \in \mathcal K} p_{\mathrm uj} \tilde{\b R}_j$, $\boldsymbol{\Psi}_{\mathrm{sum}} \triangleq \sum_{j \in \mathcal K} \boldsymbol{\Psi}_j$.
	\end{theorem}
	\begin{IEEEproof}
		See Appendix~\ref{sec:proof-UL-SINR}.
	\end{IEEEproof}
	%---------------------------------------------
	\subsection{Average Downlink SE}
	Analogously, using the UaTF scheme, the {\color{black}expression for average achievable DL sum SE} following~\eqref{rx-DL} is given by
	\begin{equation}
		\mathsf{SE_{d}}=\zeta\sum\nolimits _{k\in\mathcal{K}}\log_{2}\big(1+\gamma_{\mathrm{d}k}\big), \label{rate-DL-Def}
	\end{equation}
	where $\gamma_{\mathrm{d}k}=S_{\mathrm{d}k}/I_{\mathrm{d}k}$ with
	\begin{equation}
		S_{\mathrm{d}k}=\dfrac{\beta p_{\mathrm{b}}}{K}|\mathbb{E}\{\mathbf{u}_{k}\mathbf{f}_{k}\}|^{2}, \label{Sdk-Def}
	\end{equation}
	and 
	\begin{align}
		I_{\mathrm{d}k} = & \dfrac{\beta p_{\mathrm{b}}}{K}\mathsf{var}\big\{\mathbf{u}_{k}\mathbf{f}_{k}\big\}+\dfrac{\beta p_{\mathrm{b}}}{K}\sum_{i\in\mathcal{K}\setminus\{k\}}\mathbb{E}\big\{\big|\mathbf{u}_{k}\mathbf{f}_{i}\big|^{2}\big\} \notag \\
		& +\sum_{j\in\mathcal{K}}p_{\mathrm{u}j}\big(\sigma_{kj}^{2}+\mathbb{E}\big\{\big|\mathbf{h}_{k}\boldsymbol{\Theta}_{w_{k}}\tilde{\mathbf{h}}_{j}\big|^{2}\big\}\big)+\sigma_{\mathrm{d}}^{2}. \label{Idk-Def}
	\end{align}
	\begin{theorem} \label{thm:DL-SINR}
		Assuming that the BS uses MRT/conjugate beamforming based on the estimated CSI, i.e., $\mathbf{f}_{k}=\hat{\mathbf{u}}_{k}\herm$, closed-form expressions for $S_{\mathrm dk}$ and $I_{\mathrm dk}$ are respectively given by 
		\begin{align}
			S_{\mathrm dk} = \ \dfrac{p_{\mathrm{b}}}{\tr(\boldsymbol{\Psi}_{\mathrm{sum}})}\tr^{2}\big(\boldsymbol{\Psi}_{k}\big), \label{Sdk-Closed}
		\end{align}
		and 
		\begin{align}
			I_{\mathrm dk} = & \sum_{j\in\mathcal{K}}\bigg[ \dfrac{p_{\mathrm{b}}}{\tr(\boldsymbol{\Psi}_{\mathrm{sum}})} \tr\big(\mathbf{R}_{k}\boldsymbol{\Psi}_{j}\big) +p_{\mathrm uj}\big\{\sigma_{kj}^{2}+\delta_{h_{k}}\tilde{\delta}_{h_{j}} \notag \\
			& \times \tr\big(\b A_{w_k}\boldsymbol{\Theta}_{w_{k}}\herm\big)\big\}\bigg] -\dfrac{p_{\mathrm{b}}}{\tr(\boldsymbol{\Psi}_{\mathrm{sum}})} \tr\big(\boldsymbol{\Psi}_{k}^{2}\big)+\sigma^2_{\mathrm d}. \label{Idk-Closed} 
		\end{align}
	\end{theorem}
	\begin{IEEEproof}
		See Appendix~\ref{sec:proof-DL-SINR}.
	\end{IEEEproof}
	Note that in what follows, with a slight abuse of notation, we write
	\begin{align}
		S_{\mathrm dk} = & \ \tr^{2}\big(\boldsymbol{\Psi}_{k}\big), \label{Sdk-ClosedNew} \\
		I_{\mathrm dk} = & \ \tr\big(\mathbf{R}_{k} \boldsymbol{\Psi}_{\mathrm{sum}}\big)+\frac{\tr\big(\boldsymbol{\Psi}_{\mathrm{sum}}\big)}{p_{\mathrm b}}\sum_{j\in\mathcal{K}}\bigg[p_{\mathrm{u}j}\big\{\sigma_{kj}^{2}+\delta_{h_{k}}\tilde{\delta}_{h_{j}} \notag \\
		& \times \tr\big(\b A_{w_k}\boldsymbol{\Theta}_{w_{k}}\herm\big)\big\}\bigg]\!-\!\tr\big(\boldsymbol{\Psi}_{k}^{2}\big)+\dfrac{\sigma_{\mathrm{d}}^{2}}{p_{\mathrm{b}}}\tr(\boldsymbol{\Psi}_{\mathrm{sum}}). \label{Idk-ClosedNew} 
	\end{align}
	It can easily be seen that~\eqref{Sdk-ClosedNew} and~\eqref{Idk-ClosedNew} lead to the same $\gamma_{\mathrm dk}$ as achieved using~\eqref{Sdk-Closed} and~\eqref{Idk-Closed}. It is evident from~\eqref{Suk-Closed},~\eqref{Iuk-Closed},~\eqref{Sdk-ClosedNew}, and~\eqref{Idk-ClosedNew} that both UL and DL SE depend only on large-scale statistics in terms of correlation matrices and path-losses, which follow wide-sense stationarity and are constant for many coherence intervals~\cite{Molisch2012}. In the following, we formulate an optimization problem, which aims to maximize the average sum SE. 
	
	%---------------------------------------------
	\subsection{Problem Formulation to Maximize the Sum SE}
	With the help of~\eqref{rate-UL-Def} and~\eqref{rate-DL-Def}, the average sum SE (considering both UL and DL communications), is given by\footnote{\textcolor{black}{In the case of the HD mode, the self-interference and loop-interference terms are missing, and the pre-log factors of the SEs are different.}}
	\begin{equation}
		\mathsf{SE} = \mathsf{SE}_{\mathrm u} \!+\! \mathsf{SE}_{\mathrm d} = \zeta \sum_{k \in \mathcal K} \big\{\log_2\big(1 \!+\! \gamma_{\mathrm uk}\big) \!+\! \log_2\big(1 \!+\! \gamma_{\mathrm dk}\big)\big\}. \label{sumSE-Def}
	\end{equation}
	Note that the STARS-aided system requires the optimal design of both the receive and transmit PBMs (i.e., $\b \Theta_{\mathrm r}, \b \Theta_{\mathrm t}$) to provide the maximum performance. Notably, we account for correlated fading and imperfect CSI in an FD architecture while we assume infinite-resolution phase shifters. Therefore, the problem of finding optimal PBM designs that maximize $\mathsf{SE}$ is formulated as 
	\begin{subequations} \label{optProb}
		\begin{align}
			(\mathscr P) \qquad \underset{\boldsymbol{\theta}_{\mathrm r},\boldsymbol{\theta}_{\mathrm t}}{\maximize} & \ f\big(\boldsymbol{\theta}_{\mathrm r},\boldsymbol{\theta}_{\mathrm t}\big) \triangleq \mathsf{SE}, \label{optObj}\\
			\st & \ |\theta_{\mathrm r,n}|^2 + |\theta_{\mathrm t,n}|^2 = 1\ \forall n \in \mathcal N. \label{umc}
		\end{align}
	\end{subequations}
	In the next section, we propose a simple, yet efficient algorithm to obtain a stationary solution to~($\mathscr P$).
	%==============================
	\section{Proposed Solution}
	It can easily be noted that~($\mathscr P$) is non-convex due to the coupling between the design variables $\b \theta_{\mathrm r}$ and $\b \theta_{\mathrm t}$ in both~\eqref{optObj} and~\eqref{umc}, and the non-convex constraints in~\eqref{umc}; therefore is challenging to solve. Before proceeding further, we define the set 
	\begin{align}
		\varTheta \triangleq \{\b \theta & \ = [\b \theta_{\mathrm r}\trans, \b \theta_{\mathrm t}\trans]\trans =   [\theta_{\mathrm r,1}, \ldots, \theta_{\mathrm r,N}, \theta_{\mathrm t,1}, \ldots, \theta_{\mathrm t,N}]\trans \notag \\
		& \ \in \mathbb C^{2N\times1} \mid |\theta_{\mathrm r,n}|^2 + |\theta_{\mathrm t,n}|^2 = 1 \ \forall n \in \mathcal N\}. \label{feasibleSet}
	\end{align}
	Exploiting the fact that for a given $\hat{\b \theta}$, its Euclidean projection onto $\varTheta$ can be obtained via simple scaling, we now discuss an efficient solution using the projected gradient ascent method~\cite[Ch. 2]{Bertsekas1999}. Using~\eqref{sumSE-Def} it is straightforward to note that for $\mathrm m \in \{\mathrm r, \mathrm t\}$, we have 
	\begin{align}
		& \ \nabla_{\boldsymbol{\theta}_{\mathrm m}}f\big(\boldsymbol{\theta}_{\mathrm r},\boldsymbol{\theta}_{\mathrm t}\big) = \zeta \sum_{k \in \mathcal K} \nabla_{\boldsymbol{\theta}_{\mathrm m}} \big[\log_2(1\!+\!\gamma_{\mathrm uk}) \!+\! \log_2\big(1\!+\!\gamma_{\mathrm dk}\big)\big] \nonumber\\
		&= \zeta \log_2(e) \sum_{k \in \mathcal K} \bigg[ \frac{\nabla_{\boldsymbol{\theta}_{\mathrm m}} \gamma_{\mathrm uk}}{1+\gamma_{\mathrm uk}} + \frac{\nabla_{\boldsymbol{\theta}_{\mathrm m}}\gamma_{\mathrm{dk}}}{1+\gamma_{\mathrm dk}}\bigg] \notag \\
		&= \zeta \log_2(e) \sum_{k \in \mathcal K} \bigg[ \frac{1}{(1+\gamma_{\mathrm uk}) I_{\mathrm uk}^2} \big(I_{\mathrm uk} \nabla_{\boldsymbol{\theta}_{\mathrm m}} S_{\mathrm uk} \!-\! S_{\mathrm uk} \nabla_{\boldsymbol{\theta}_{\mathrm m}} I_{\mathrm uk}\big) \nonumber\\
		&\hspace{1cm}+ \frac{1}{(1+\gamma_{\mathrm dk}) I_{\mathrm dk}^2} \big(I_{\mathrm dk} \nabla_{\boldsymbol{\theta}_{\mathrm m}} S_{\mathrm dk} - S_{\mathrm dk} \nabla_{\boldsymbol{\theta}_{\mathrm m}} I_{\mathrm dk}\big) \bigg]. \label{grad-f-R-Def}
	\end{align}
	In the theorem shown below, we obtain the expression for different gradient terms involved in~\eqref{grad-f-R-Def}, \textcolor{black}{which are obtained in closed forms in terms of products of covariance matrices.}
	\begin{theorem} \label{thm:gradClosed}
		Considering $\mathrm m \in \{\mathrm r, \mathrm t\}$, closed-form expressions for $\nabla_{\boldsymbol{\theta}_{\mathrm m}}S_{\mathrm uk}$, $\nabla_{\boldsymbol{\theta}_{\mathrm m}}S_{\mathrm dk}$, $\nabla_{\boldsymbol{\theta}_{\mathrm m}}I_{\mathrm uk}$, and $\nabla_{\boldsymbol{\theta}_{\mathrm m}}I_{\mathrm dk}$ are given by~\eqref{gradSuk-thetaTR-Closed} -- \eqref{grad-thetaT-Idk-Closed}: 
		\begin{equation}
			\nabla_{\boldsymbol{\theta}_{\mathrm m}} S_{\mathrm uk} = \begin{cases} \tilde \nu_k \diag(\mathbf A_{\mathrm r}), & \text{if } \mathrm m = \mathrm r \\ \boldsymbol 0, & \text{otherwise } \end{cases}, \label{gradSuk-thetaTR-Closed}
		\end{equation}
		\begin{equation}
			\nabla_{\boldsymbol \theta_{\mathrm r}} S_{\mathrm dk} = \begin{cases} \nu_k \diag\big(\mathbf A_{\mathrm r}\big), & \text{if } w_k = \mathrm r \\ \boldsymbol 0, & \text{otherwise} \end{cases}, \label{gradSdk-thetaR-Closed}
		\end{equation}
		\begin{equation}
			\nabla_{\boldsymbol \theta_{\mathrm t}} S_{\mathrm dk} = \begin{cases} \nu_k \diag\big(\mathbf A_{\mathrm t}\big), & \text{if } w_k = \mathrm t \\ \boldsymbol 0, & \text{otherwise} \end{cases}, \label{gradSdk-thetaT-Closed} 
		\end{equation}
		\begin{figure*}
			\begin{align}
				\nabla_{\boldsymbol{\theta}_{\mathrm{r}}}I_{\mathrm{u}k} = & \Big[\tilde{\delta}_{g}\tilde{\delta}_{h_{k}}\tr\big\{{\tilde{\mathbf{R}}_{\mathrm{b}}\big(\tilde{\mathbf{B}}_{k1}+\tilde{\mathbf{B}}_{k2}-\tilde{\mathbf{B}}_{k3}+\sigma_{\mathrm{u}}^{2}\tilde{\mathbf{C}}_{k}\big)}\big\} \notag \\ & \ +\tilde{\delta}_{g}\big(\sum_{j\in\mathcal{K}}\tilde{\delta}_{h_{j}} p_{\mathrm uj}\big)\tr\big\{{\tilde{\mathbf{R}}_{\mathrm{b}}\tilde{\boldsymbol{\Psi}}_{k}}\big\}+\delta_{g}\sum_{{i\in\mathcal{K}_{\mathrm{r}}}}\delta_{h_{i}}\tr\big\{{\mathbf{R}_{\mathrm{b}}\tilde{\mathbf{L}}_{ki}}\big\}+\tilde{\chi}_{k}\Big]\diag\big(\mathbf{A}_{\mathrm{r}}\big), \label{grad-thetaR-Iuk-Closed}
			\end{align}	
		\end{figure*}%
		\begin{equation}
			\nabla_{\boldsymbol{\theta}_{\mathrm{t}}}I_{\mathrm{u}k}=\Big[\delta_{g}\sum \nolimits_{{i\in\mathcal{K}_{\mathrm{t}}}}\delta_{h_{i}}\tr\big\{{\mathbf{R}_{\mathrm{b}}\tilde{\mathbf{L}}_{ki}}\big\}\Big]\diag\big(\mathbf{A}_{\mathrm{t}}\big), \label{grad-thetaT-Iuk-Closed}		
		\end{equation}%
		\begin{figure*}
			\begin{align}
				\nabla_{\boldsymbol{\theta}_{\mathrm{r}}}I_{\mathrm{d}k} & =\begin{cases}
					\Big[\delta_{g}\delta_{h_{k}}\tr\big\{\mathbf{R}_{\mathrm{b}}\big(\boldsymbol{\Psi}_{\mathrm{sum}}-\mathbf{B}_{k}\big)\big\} +\delta_{g}\sum_{i \in\mathcal{K}_{\mathrm{r}}}\delta_{h_{i}}\tr\big\{\mathbf{R}_{\mathrm{b}}\big(\mathbf{L}_{ki}+\chi_{k1}\b C_i \big)\big\}+\chi_{k2}\Big]\diag\big(\mathbf{A}_{\mathrm{r}}\big), & \text{if }w_{k}=\mathrm{r}\\
					\Big[\delta_{g}\sum_{i \in\mathcal{K}_{\mathrm{r}}}\delta_{h_i}\tr\big\{\mathbf{R}_{\mathrm{b}}\big(\mathbf{L}_{ki}+\chi_{k1}\b C_i \big)\big\}\Big]\diag\big(\mathbf{A}_{\mathrm{r}}\big), & \text{otherwise}
				\end{cases}, \label{grad-thetaR-Idk-Closed}
			\end{align}
		\end{figure*}%
		\begin{figure*}
			\begin{align}
				\nabla_{\boldsymbol{\theta}_{\mathrm{t}}}I_{\mathrm{d}k} & =\begin{cases}\Big[\delta_{g}\sum_{i \in\mathcal{K}_{\mathrm{t}}}\delta_{h_i}\tr\big\{\mathbf{R}_{\mathrm{b}}\big(\mathbf{L}_{ki}+\chi_{k1}\b C_{\imath} \big)\big\}\Big]\diag\big(\mathbf{A}_{\mathrm{t}}\big), & \text{if }w_{k}=\mathrm{r} \\
					\Big[\delta_{g}\delta_{h_{k}}\tr\big\{\mathbf{R}_{\mathrm{b}}\big(\boldsymbol{\Psi}_{\mathrm{sum}}-\mathbf{B}_{k}\big)\big\} +\delta_{g}\sum_{i \in\mathcal{K}_{\mathrm{t}}}\delta_{h_i}\tr\big\{\mathbf{R}_{\mathrm{b}}\big(\mathbf{L}_{ki}+\chi_{k1}\b C_i \big)\big\}+\chi_{k2}\Big]\diag\big(\mathbf{A}_{\mathrm{t}}\big), & \text{otherwise}
				\end{cases}, \label{grad-thetaT-Idk-Closed}
			\end{align}	
		\end{figure*}%
		
		\noindent where 
		$\tilde \nu_k \triangleq 2 \tilde \delta_g \tilde \delta_{h_k} p_{\mathrm uk} \tr \big( \tilde{\boldsymbol{\Psi}}_k \big) \tr \big( \tilde{\mathbf R}_{\mathrm{b}} \tilde{\mathbf C}_k  \big),$
		$\tilde{\mathbf C}_k \triangleq \tilde{\mathbf Q}_k \tilde{\mathbf R}_k - \tilde{\mathbf Q}_k \tilde{\mathbf R}_k^2 \tilde{\mathbf Q}_k + \tilde{\mathbf R}_k \tilde{\mathbf Q}_k,$
		$\nu_k \triangleq 2 \delta_g \delta_{h_k} \tr \big( \boldsymbol{\Psi}_k \big) \tr \big(\mathbf R_{\mathrm{b}} \mathbf C_k  \big),$
		$\mathbf C_k \triangleq \mathbf Q_k \mathbf R_k - \mathbf Q_k \mathbf R_k^2 \mathbf Q_k + \mathbf R_k \mathbf Q_k,$
		$\tilde{\mathbf{B}}_{k1} \triangleq \tilde{\mathbf{Q}}_{k}\tilde{\mathbf{R}}_{k}\tilde{\mathbf{R}}_{\mathrm{sum}}-\tilde{\mathbf{Q}}_{k}\tilde{\mathbf{R}}_{k}\tilde{\mathbf{R}}_{\mathrm{sum}}\tilde{\mathbf{R}}_{k}\tilde{\mathbf{Q}}_{k}+\tilde{\mathbf{R}}_{\mathrm{sum}}\tilde{\mathbf{R}}_{k}\tilde{\mathbf{Q}}_{k},$
		$\varpi \triangleq \frac{p_{\mathrm{b}}}{\tr\big(\boldsymbol{\Psi}_{\mathrm{sum}}\big)}\tr\big(\mathbf{R}_{\mathrm{b}}\boldsymbol{\Psi}_{\mathrm{sum}}\big)\big\{\delta_{g}\tilde{\delta}_{g}\tr\big(\mathbf{A}_{\mathrm{r}}\boldsymbol{\Theta}_{\mathrm{r}}\herm\big)+\sigma_{\mathrm{L}}^{2}\big\},$ 
		$\tilde{\boldsymbol{\Xi}}_{k} \triangleq \frac{p_{\mathrm{b}}}{\tr\big(\boldsymbol{\Psi}_{\mathrm{sum}}\big)}\tr\big(\tilde{\boldsymbol{\Psi}}_{k}\tilde{\mathbf{R}}_{\mathrm{b}}\big)\big\{\delta_{g}\tilde{\delta}_{g}\tr\big(\mathbf{A}_{\mathrm{r}}\boldsymbol{\Theta}_{\mathrm{r}}\herm\big)+\sigma_{\mathrm{L}}^{2}\big\}\big[\mathbf{R}_{\mathrm{b}}-\frac{\tr\big(\mathbf{R}_{\mathrm{b}}\boldsymbol{\Psi}_{\mathrm{sum}}\big)}{\tr\big(\boldsymbol{\Psi}_{\mathrm{sum}}\big)}\big],$
		$\tilde{\chi}_{k}\triangleq\frac{p_{\mathrm{b}}\delta_{g}\tilde{\delta}_{g}}{\tr\big(\boldsymbol{\Psi}_{\mathrm{sum}}\big)}\tr\big(\tilde{\boldsymbol{\Psi}}_{k}\tilde{\mathbf{R}}_{\mathrm{b}}\big)\tr\big(\mathbf{R}_{\mathrm{b}}\boldsymbol{\Psi}_{\mathrm{sum}}\big),$ 
		$\tilde{\mathbf{B}}_{k2} \triangleq \varpi \big(\tilde{\mathbf{Q}}_{k}\tilde{\mathbf{R}}_{k}\tilde{\mathbf{R}}_{\mathrm{b}}-\tilde{\mathbf{Q}}_{k}\tilde{\mathbf{R}}_{k}\tilde{\mathbf{R}}_{\mathrm{b}}\tilde{\mathbf{R}}_{k}\tilde{\mathbf{Q}}_{k}+\tilde{\mathbf{R}}_{\mathrm{b}}\tilde{\mathbf{R}}_{k}\tilde{\mathbf{Q}}_{k}\big),$ 
		$\tilde{\mathbf{L}}_{kj} \triangleq \mathbf{Q}_{j}\mathbf{R}_{j}\tilde{\boldsymbol{\Xi}}_{k}-\mathbf{Q}_{j}\mathbf{R}_{j}\tilde{\boldsymbol{\Xi}}_{k}\mathbf{R}_{j}\mathbf{Q}_{j}+\tilde{\boldsymbol{\Xi}}_{k}\mathbf{R}_{j}\mathbf{Q}_{j},$
		$\tilde{\mathbf{B}}_{k3}=2p_{\mathrm{u}k}\big(\tilde{\mathbf{Q}}_{k}\tilde{\mathbf{R}}_{k}\tilde{\boldsymbol{\Psi}}_{k}-\tilde{\mathbf{Q}}_{k}\tilde{\mathbf{R}}_{k}\tilde{\boldsymbol{\Psi}}_{k}\tilde{\mathbf{R}}_{k}\tilde{\mathbf{Q}}_{k}+\tilde{\boldsymbol{\Psi}}_{k}\tilde{\mathbf{R}}_{k}\tilde{\mathbf{Q}}_{k}\big),$
		$\mathbf{L}_{kj}=\mathbf{Q}_{j}\mathbf{R}_{j}\mathbf{R}_{k}-\mathbf{Q}_{j}\mathbf{R}_{j}\mathbf{R}_{k}\mathbf{R}_{j}\mathbf{Q}_{j}+\mathbf{R}_{k}\mathbf{R}_{j}\mathbf{Q}_{j},$  
		$\mathbf{B}_{k} \triangleq 2\big(\mathbf{Q}_{k}\mathbf{R}_{k}\boldsymbol{\Psi}_{k}-\mathbf{Q}_{k}\mathbf{R}_{k}\boldsymbol{\Psi}_{k}\mathbf{R}_{k}\mathbf{Q}_{k}+\boldsymbol{\Psi}_{k}\mathbf{R}_{k}\mathbf{Q}_{k}\big),$
		$\chi_{k1} \triangleq \frac{1}{p_{\mathrm{b}}}\sum_{j\in\mathcal{K}}p_{\mathrm{u}j}\Big\{\sigma_{kj}^{2}+\delta_{h_{k}}\tilde{\delta}_{h_{j}}\tr\big(\mathbf{A}_{w_{k}}\boldsymbol{\Theta}_{w_{k}}\herm\big)\Big\}+\frac{\sigma_{\mathrm{d}}^{2}}{p_{\mathrm{b}}},$  
		and
		$\chi_{k2} \triangleq \frac{\delta_{h_{k}}}{p_{\mathrm{b}}}\tr\big(\boldsymbol{\Psi}_{\mathrm{sum}}\big)\sum_{j\in\mathcal{K}}p_{\mathrm{u}j}\tilde{\delta}_{h_{j}}.$
	\end{theorem}
	\begin{IEEEproof}
		See Appendix~\ref{sec:proof-gradClosed}.
	\end{IEEEproof}

	\begin{algorithm}[t]
		\caption{The Proposed ProGrAM to solve~($\mathscr P$).} \label{algo}
		\KwIn{$\b \theta_{\mathrm r}^{(0)}$, $\b \theta_{\mathrm t}^{(0)}$, $\mu_1 > 0$, $\epsilon$}
		\KwOut{$\b \theta_{\mathrm r}^{\mathrm{opt}}$, $\b \theta_{\mathrm t}^{\mathrm{opt}}$}
		$\ell \leftarrow 1$\;
		\Repeat{$\big\{f\big(\b \theta_{\mathrm r}^{(\ell)}, \b \theta_{\mathrm t}^{(\ell)})-f\big(\b \theta_{\mathrm r}^{(\ell -1)}, \b \theta_{\mathrm t}^{(\ell -1)})\big\}/f\big(\b \theta_{\mathrm r}^{(\ell -1)}, \b \theta_{\mathrm t}^{(\ell -1)}) \geq \epsilon$}
		{
			\tcc{Find the gradients}
			Obtain $\nabla_{\b \theta_{\mathrm r}} f\big(\b \theta_{\mathrm r}^{(\ell -1)}, \b \theta_{\mathrm t}^{(\ell -1)})$ and $\nabla_{\b \theta_{\mathrm t}} f\big(\b \theta_{\mathrm r}^{(\ell -1)}, \b \theta_{\mathrm t}^{(\ell -1)})$ using~\eqref{grad-f-R-Def} -- \eqref{grad-thetaT-Idk-Closed} \;
			%
			%\tcc{Project onto the feasible set}
			\tcc{Move $\boldsymbol \theta_{\mathrm r}$ along $\nabla_{\b \theta_{\mathrm r}}f(\cdot)$ direction}
			$\hat{\b \theta}_{\mathrm r}^{(\ell )} \leftarrow \b \theta_{\mathrm r}^{(\ell -1)} + \mu_\ell \nabla_{\b \theta_{\mathrm r}} f\big(\b \theta_{\mathrm r}^{(\ell -1)}, \b \theta_{\mathrm t}^{(\ell -1)})$\;  %\tcp*{move along $\nabla_{\b \theta_{\mathrm r}}f(\cdot)$ direction}
			\tcc{Move $\boldsymbol \theta_{\mathrm t}$ along $\nabla_{\b \theta_{\mathrm t}}f(\cdot)$ direction}
			$\hat{\b \theta}_{\mathrm t}^{(\ell )} \leftarrow \b \theta_{\mathrm t}^{(\ell -1)} + \mu_\ell \nabla_{\b \theta_{\mathrm t}} f\big(\b \theta_{\mathrm r}^{(\ell -1)}, \b \theta_{\mathrm t}^{(\ell -1)})$\; %\tcp*{move along $\nabla_{\b \theta_{\mathrm t}}f(\cdot)$ direction}
			\tcc{concatenate $\hat{\b \theta}_{\mathrm r}^{(\ell )}$ and $\hat{\b \theta}_{\mathrm t}^{(\ell )}$}
			$\hat{\b \theta}^{\ell} \triangleq \big[\big(\hat{\b \theta}_{\mathrm r}^{(\ell)}\big)\trans, \big(\hat{\b \theta}_{\mathrm t}^{(\ell)}\big)\trans\big]\trans$ \;
			\tcc{Project onto feasible $\varTheta$}
			$\b \theta^{(\ell)} \leftarrow \Pi_{\varTheta}\big\{\hat{\b \theta}^{\ell}\big\}$ \;
			$\mu_{\ell+1} \leftarrow \mu_\ell$\;
			$\ell \leftarrow \ell+1$ \tcp*{update iteration index}
		}
		$\b \theta_{\mathrm r}^{\mathrm{opt}} \leftarrow \b \theta_{\mathrm r}^{(\ell)}$, $\b \theta_{\mathrm t}^{\mathrm{opt}} \leftarrow \b \theta_{\mathrm t}^{(\ell)}$\;
	\end{algorithm}
	
	With \textcolor{black}{ the background above}, we are now in a position to discuss ProGrAM as shown in~\textbf{Algorithm~\ref{algo}}. In this algorithm, in the $\ell^{\text{th}}$ iteration with given the initial points $\b \theta_{\mathrm r}^{(\ell)}$, $\b \theta_{\mathrm t}^{(\ell)}$, we first obtain $\nabla_{\b \theta_{\mathrm r}} f\big(\b \theta_{\mathrm r}^{(\ell-1)}, \b \theta_{\mathrm t}^{(\ell-1)})$ and $\nabla_{\b \theta_{\mathrm t}} f\big(\b \theta_{\mathrm r}^{(\ell-1)}, \b \theta_{\mathrm t}^{(\ell-1)})$ (see line~3). After obtaining the gradients, we ascent in the gradient direction to obtain $\hat{\b \theta}_{\mathrm r}^{(\ell)}$ and $\hat{\b \theta}_{\mathrm t}^{(\ell)}$ (see lines~4 and~5). Next, we project the new vector $\hat{\b \theta}^{\ell}$ (see line~6) onto the set $\varTheta$ (as shown in line~7). Note that for a given $\hat{\boldsymbol{\theta}}^{(\ell)}=\big[\hat{\theta}_{\mathrm{r},1}^{(\ell)},\hat{\theta}_{\mathrm{r},2}^{(\ell)},\ldots,\hat{\theta}_{\mathrm{r},N}^{(\ell)},\hat{\theta}_{\mathrm{t},1}^{(\ell)},\hat{\theta}_{\mathrm{t},2}^{(\ell)},\ldots,\hat{\theta}_{\mathrm{t},N}^{(\ell)}\big]\trans$,
	its projection onto $\varTheta$, i.e., $\Pi_{\varTheta}\big(\hat{\boldsymbol{\theta}}^{(\ell)}\big)$,
	is given by $\boldsymbol{\theta}^{(\ell)}=\big[\theta_{\mathrm{r},1}^{(\ell)},\theta_{\mathrm{r},2}^{(\ell)},\ldots,\theta_{\mathrm{r},N}^{(\ell)},\theta_{\mathrm{t},1}^{(\ell)},\theta_{\mathrm{t},2}^{(\ell)},\ldots,\theta_{\mathrm{t},N}^{(\ell)}\big]\trans$,
	where $\big\{\theta_{\mathrm{r},n}^{(\ell)},\theta_{\mathrm{t},n}^{(\ell)}\big\}$ is given by~\eqref{eq:projTheta}.
	\begin{figure*}
		\begin{equation}
			\big\{\theta_{\mathrm{r},n}^{(\ell)},\theta_{\mathrm{t},n}^{(\ell)}\big\}\!\! =\!\! 
			\begin{cases} \!\!
				\bigg\{\!\!\frac{\hat{\theta}_{\mathrm{r},n}^{(\ell)}}{\sqrt{\big(\hat{\theta}_{\mathrm{r},n}^{(\ell)}\big)^2+\big(\hat{\theta}_{\mathrm{t},n}^{(\ell)}\big)^2}},\frac{\hat{\theta}_{\mathrm{t},n}^{(\ell)}}{\sqrt{\big(\hat{\theta}_{\mathrm{r},n}^{(\ell)}\big)^2+\big(\hat{\theta}_{\mathrm{t},n}^{(\ell)}\big)^2}}\bigg\}, &\!\!\! \mathrm{if}\ \sqrt{\big(\hat{\theta}_{\mathrm{r},n}^{(\ell)}\big)^2+\big(\hat{\theta}_{\mathrm{t},n}^{(\ell)}\big)^2} \neq 0\\
				\big\{\sqrt{0.5}\exp(j\phi),\sqrt{0.5}\exp(j\phi)\big\},\phi\in[0,2\pi[, & \!\!\mathrm{otherwise}
			\end{cases}, \forall n\in\mathcal{N}. \label{eq:projTheta}
		\end{equation}
		\hrulefill 
	\end{figure*}
	It is noteworthy that choosing an appropriate value of step-size $\mu_{\ell}$ is critical to control the speed of convergence for the algorithm, which can efficiently be obtained by using the Barzilai-Borwein (BB) rule (see~\cite{StepSize}, and the discussion therein).

	\textcolor{black}{Regarding the complexity (number of complex multiplications)  per iteration of the proposed algorithm, we use the big-$ \mathcal{O} $ notation, which is quite meaningful for large $ M_{\mathrm T}$ and $ M_{\mathrm R}$ used by the proposed architecture. In particular, we find that the complexity of  the objective function is $ \mathcal{O}(K(N (M_{\mathrm T}^{3}+M_{\mathrm R}^{3})+N^{2} (M_{\mathrm T}+M_{\mathrm R}) N^{3}+N^{3})) $. Next, by focusing on  the complexity of the computation of the gradients, we observe that each of the   gradients requires a similar complexity to the objective function.}

	%==============================
	\section{Numerical Results and Discussion}
\begin{figure}
		\centering
		\includegraphics[width=0.8\columnwidth]{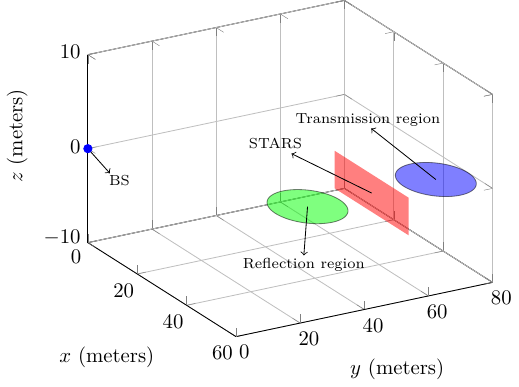}
		\caption{Simulation setup}
		\label{setup}
	\end{figure}

	In this section, \textcolor{black}{we present extensive numerical   analytical results and Monte-Carlo (MC) simulations with $ 10^{3} $ independent channel realizations }  \textcolor{black}{of \eqref{sumSE-Def}} to obtain deeper insights into the system performance and its dependence on various system parameters of interest. For this purpose, we assume that the locations of the BS and STARS in a 2D Euclidean coordinate system are respectively given by $(x_{\mathrm B}, y_{\mathrm B}) = (0,0)$ and $(x_{\mathrm S}, y_{\mathrm S}) = (50,10)$, respectively, all in meter units. \textcolor{black}{The UEs in the reflection regime are assumed to be distributed on two setups. The first setup assumes a straight line between $(x_{\mathrm S}-0.5d_0, y_{\mathrm S} - 0.5d_0)$ and $(x_{\mathrm S}+0.5d_0, y_{\mathrm S} - 0.5d_0)$, with the distance between two adjacent UEs being constant and $d_0 = 20$~m. Similarly, the UEs in the transmission region are assumed to be located at an equal distance from the adjacent UE on the straight line between $(x_{\mathrm S} - 0.5d_0, y_{\mathrm S} + 0.5d_0)$ and $(x_{\mathrm S} + 0.5d_0, y_{\mathrm S} + 0.5d_0)$. This setup is used in figures below unless otherwise stated. ii) The second setup assumes  randomly located users on  circular regions of radius $10~\mathrm{m}$. The second setup is more general and is illustrated in Fig. \ref{setup}}.   The correlation matrices $\b R_{\mathrm{b}}$ and $\b R_{\mathrm{s}} $ are obtained by following~ \cite{Hoydis2013} and~\cite{Bjoernson2020}, respectively. We assume that the size of each of the STARS elements is $d_{\mathrm H} = d_{\mathrm V}$. The distance-based path loss between the transmit array of the BS and the STARS is given by $\delta_g = d_{\mathrm H} d_{\mathrm V} \wp^{-\alpha}$, where $\wp$ is the distance between the BS and the STARS. A similar modeling is applied for $\tilde{\delta}_g$, $\delta_{h_k}$ and $\tilde{\delta}_{h_k}$. Unless stated otherwise, in this section we consider $M_{\mathrm T} = M_{\mathrm R} = 128$, $N = 144$, $\lambda = 0.1$~m, $K_{\mathrm r} = K_{\mathrm t} = 2$, $\tau_{\mathrm c} = 200$, $\tau_{\mathrm{up}} = \tau_{\mathrm{dp}} = K$, $p_{\mathrm b} = 30$~dBm, $p_{\mathrm uk} = 15$~dBm ($\forall k \in \mathcal K$), $p_{\mathrm{train}} = 15$~dBm, $\sigma^2 = \sigma^2_{\mathrm b} = \sigma^2_{\mathrm d} = -94$~dBm, $\sigma^2_{\mathrm L} = 0$~dB w.r.t $\sigma^2$, $\alpha = 2.6$, $\mu_1 = 500$, and $\epsilon = 10^{-5}$.\textcolor{black}{ Note that $\lambda = 0.1$~m corresponds to a frequency of $3$ GHz, i.e., we focus sub-6 GHz region.} Moreover, the value of $\sigma^2_{kj}$ w.r.t. $\sigma^2$ is 
	\begin{equation*}
		\sigma^2_{kj} = \begin{cases} 0~\text{dB}, & \text{if } w_j = w_k \\ 0, & \text{otherwise} \end{cases}.
	\end{equation*}
	\begin{figure*}[t]
		\begin{minipage}{0.32\textwidth}
			\centering
			\includegraphics[width=\linewidth]{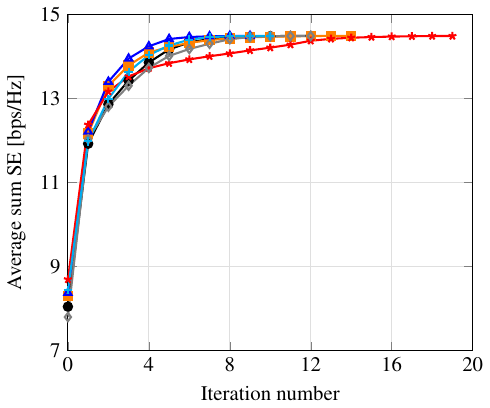}
			\caption{Impact of random initial points ($\boldsymbol{\theta}_{\mathrm r}^{(0)}, \boldsymbol{\theta}_{\mathrm t}^{(0)}$) on the convergence of the proposed ProGrAM.}
			\label{fig:convSeq}
		\end{minipage}%
		\hfill
		\begin{minipage}{0.32\textwidth}
			\centering
			\includegraphics[width=\linewidth]{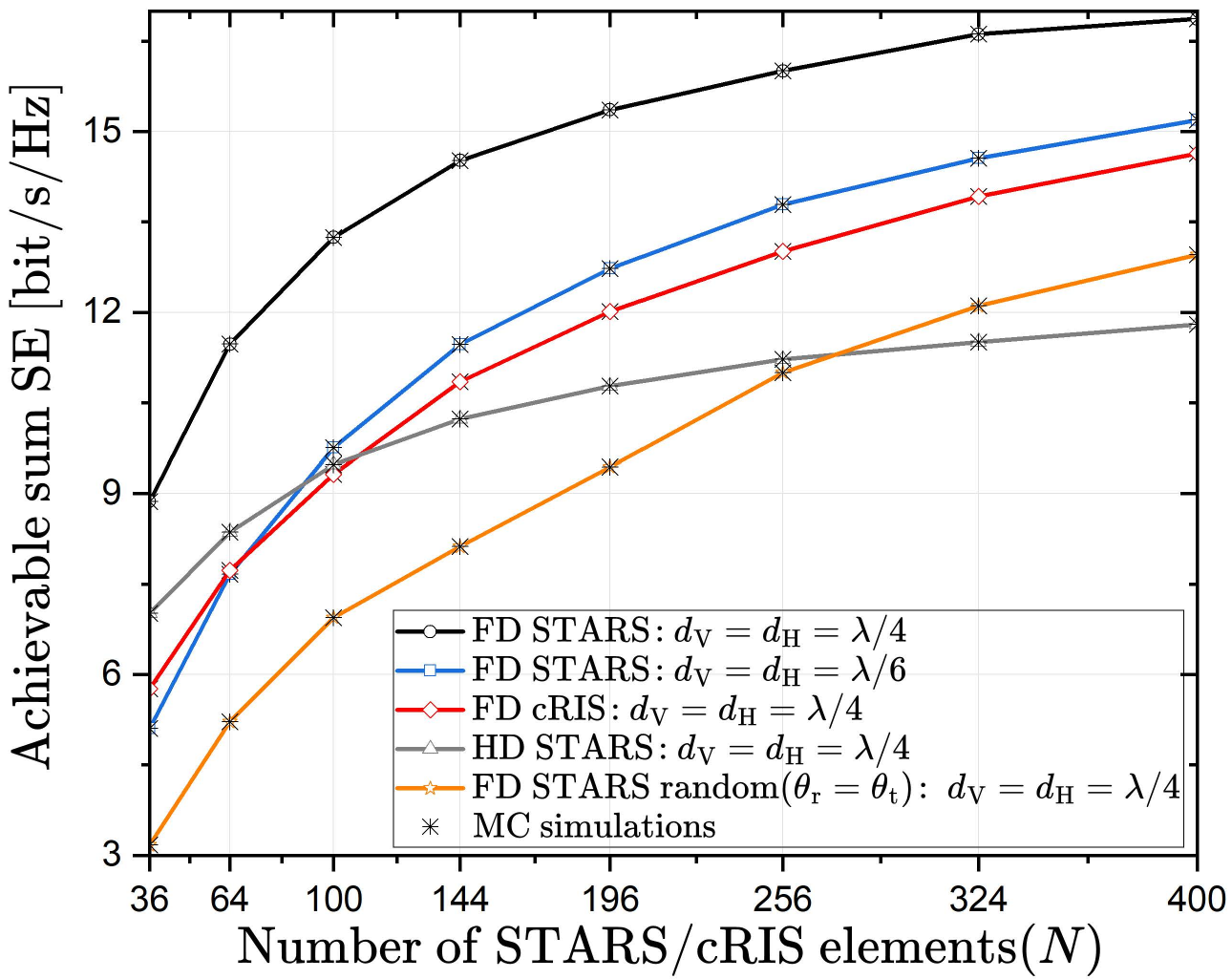}
			\caption{Impact of the number of STARS elements on the average sum SE.}
			\label{fig:rateN}
		\end{minipage}%
		\hfill 
		\begin{minipage}{0.32\textwidth}
			\centering
			\includegraphics[width=\linewidth]{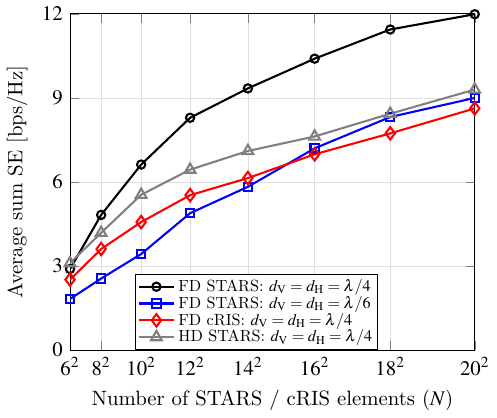}
			\caption{Impact of the BS transmit power on the FD system performance for fixed $p_{\mathrm u} = 15$~dBm.}
			\label{fig:rateN1}
		\end{minipage}%
	\end{figure*}
	The impact of various initial points on the convergence of the ProGrAM in obtaining a stationary solution to problem $(\mathscr P)$ is depicted in Fig.~\ref{fig:convSeq}. Specifically, we have plotted the average sum SE  versus the iteration number returned by \textbf{Algorithm~\ref{algo}} from 5 different randomly generated initial $\big(\b \theta_{\mathrm r}^{(0)}, \b \theta_{\mathrm t}^{(0)}\big)$. \textcolor{black}{Note that we have checked the convergence for many different initial points but we have selected 5 points for the clarity of the figure.} It is observed that the proposed optimized solution is insensitive to the initialization and the algorithm converges rapidly within a small number of iterations. 
	
	In Fig.~\ref{fig:rateN}, we show the dependence of the average sum SE versus the number of elements in the STARS. As the number of elements increases, the STARS can form narrower beams towards the UEs and/or BS by reducing interference, thus resulting in an increase in the sum SE. It is noteworthy that as the size of the STARS elements decreases (from $\lambda/4$ to $\lambda/6$), the distance between the STARS elements also decreases resulting in a stronger correlation at the STARS. This increased correlation at the STARS has a detrimental effect on system performance, as evident from the figure. Furthermore, we compare the performance of the STARS-enabled FD system with that of a corresponding STARS-enabled HD system and a cRIS-enabled FD system. \textcolor{black}{In the case of the FD protocol, the BS transmits $p_{\mathrm b}$ units of power and all of the users transmit $p_{\mathrm u}$ units of power. Hence the total power used in one channel use is $(p_{\mathrm b} + K p_{\mathrm u})$. On the other hand, in the case of HD protocol, the BS transmits using $p_{\mathrm b}$ units of power for half channel use, and in the next half of channel use, all the users simultaneously transmit the signal using $p_{\mathrm u}$ units of power. Therefore, the total power used in this case is $(p_{\mathrm b} + K p_{\mathrm u})$ units of power; this indeed is the same as that used in the FD protocol. The difference between the two protocols lies in computing the spectral efficiency. In the case of the HD protocol, a factor of $0.5$ appears due to half channel use.  Moreover,  the cRIS system is assumed to be composed of a transmit-only RIS and a reflect-only RIS with $N/2$ elements each. Regarding the optimization of the cRIS, it changes in terms of the objective function and the  constraints. In particular, we have two optimization problems, one for the transmit-only RIS and one for the reflect-only RIS.}  \textcolor{black}{The advantages of the STARS-enabled FD system over both the STARS-enabled HD system and cRIS-enabled FD systems are clearly seen in the figure.} Additionally, the significantly inferior performance of the STARS-enabled FD system with random PBM, as shown in the figure, reveals the benefit of obtaining the optimal PBM. \textcolor{black}{Compared to Fig.~\ref{fig:rateN} corresponding to the first simulation setup with users in straight lines, in Fig.~\ref{fig:rateN1}, we depict the impact of random realizations of the users' positions based on the second simulation setup. We observe that the rate is lower since the inter-user interference is larger due to decreased inter-user distance compared to the first geometry. }
	
	\begin{figure*}[t]
		\begin{minipage}{0.45\textwidth}
			\centering
			\includegraphics[width=0.7\linewidth]{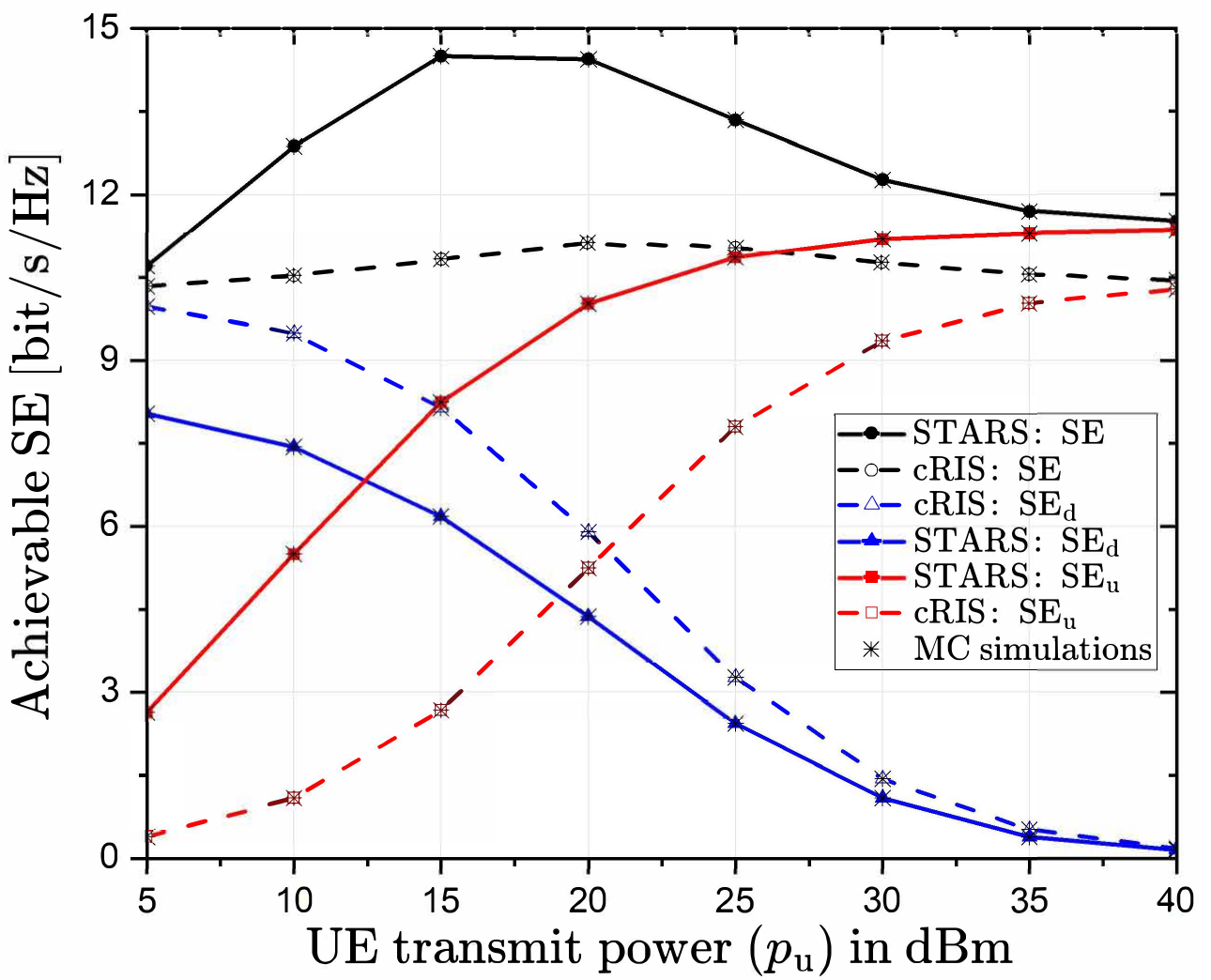}
			\caption{Impact of UEs' transmit power on the FD system performance for fixed $p_{\mathrm b} = 30$~dBm.}
			\label{fig:ratePu}
		\end{minipage}%
		\hfill 
		\begin{minipage}{0.45\textwidth}
			\centering
			\includegraphics[width=0.7\linewidth]{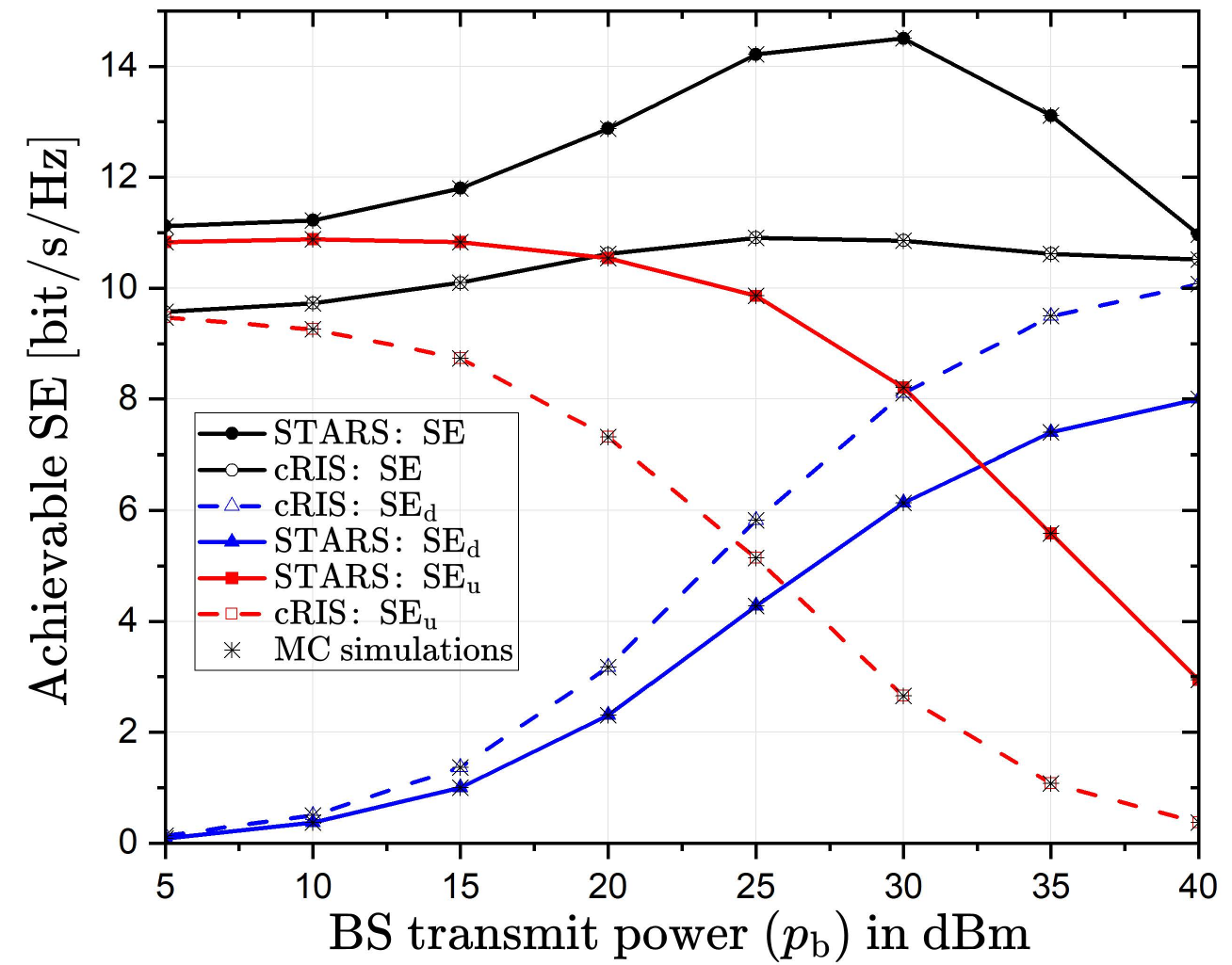}
			\caption{Impact of random realizations of the users' positions based on the second simulation setup}
			\label{fig:ratePb}
		\end{minipage}%
	\end{figure*}

	\begin{figure*}[t]
		\begin{minipage}{0.45\textwidth}
			\centering
			\includegraphics[width=0.7\linewidth]{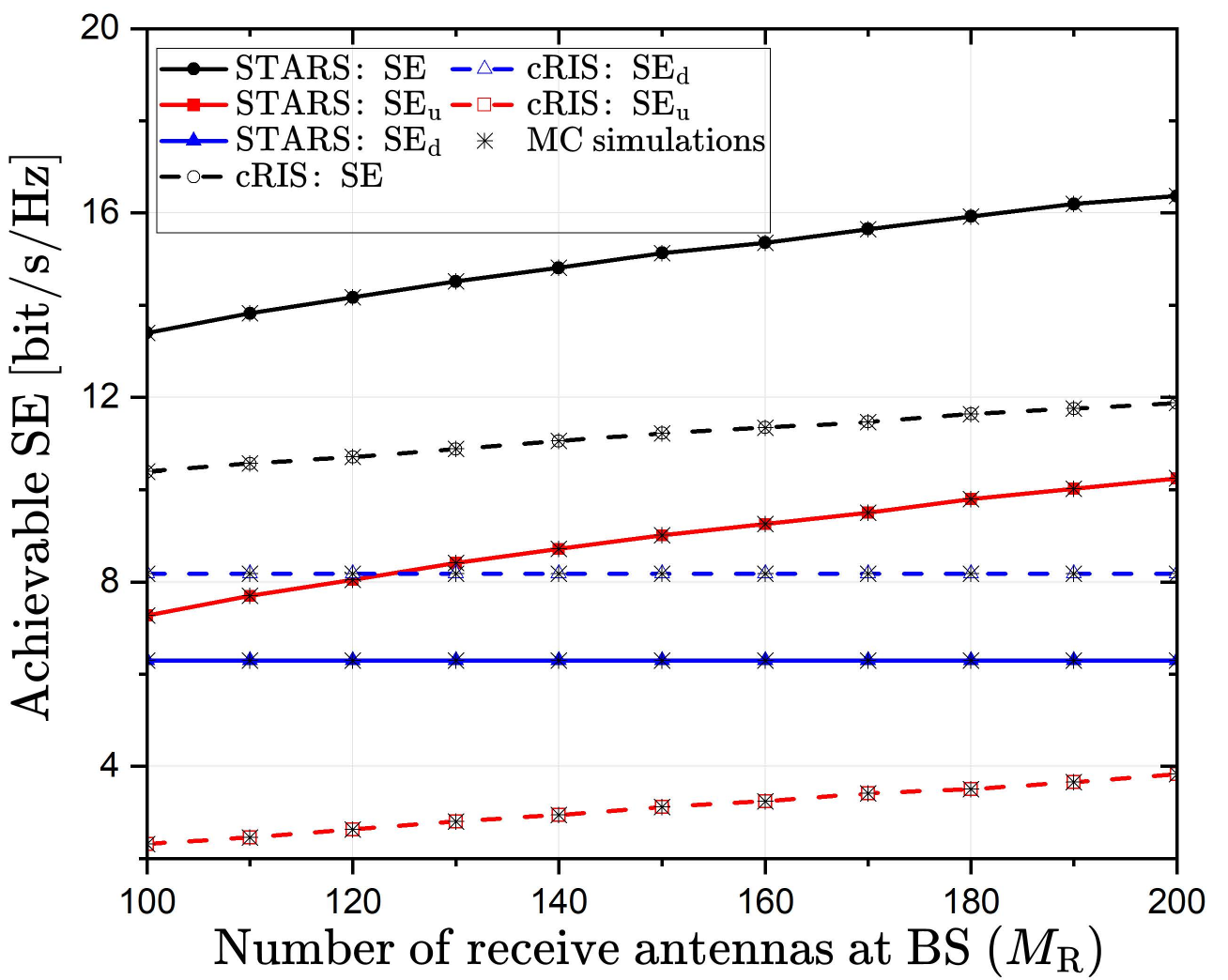}
			\caption{Impact of the number of receive antennas at the BS on the FD system performance at fixed $M_{\mathrm T} = 128$.}
			\label{fig:rateMr}
		\end{minipage}%
		\hfill
		\begin{minipage}{0.45\textwidth}
			\centering
			\includegraphics[width=0.7\linewidth]{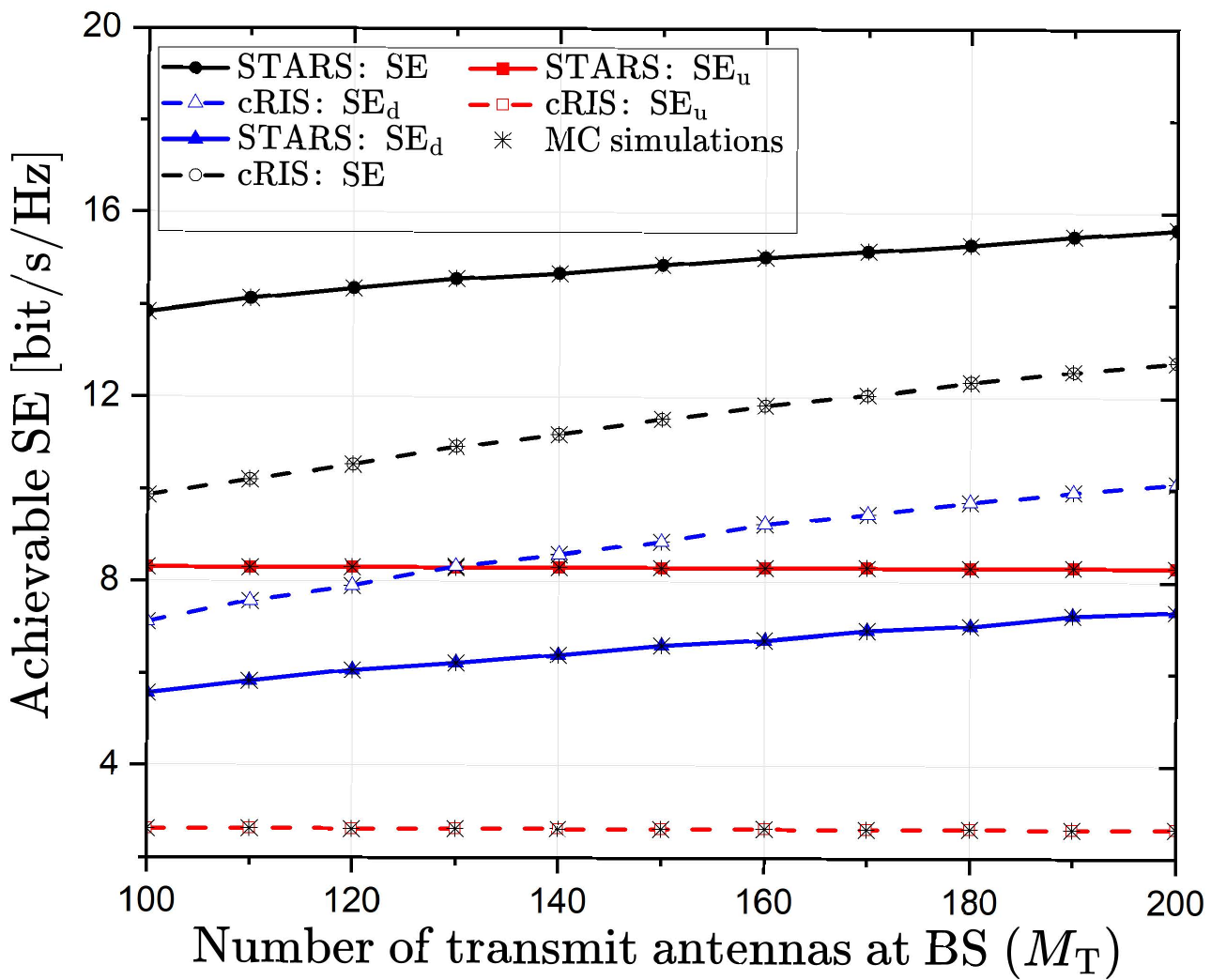}
			\caption{Impact of the number of transmit antennas at the BS on the FD system performance at fixed $M_{\mathrm R} = 128$.}
			\label{fig:rateMt}
		\end{minipage}
	\end{figure*}
	
	The variation of the SE of the STARS- and cRIS-enabled FD system w.r.t. $p_{\mathrm b}$ is shown in Fig.~\ref{fig:ratePb}, while other quantities in the system are fixed. Note that an increase in $p_{\mathrm b}$ increases the SINR at the DL UEs (see~\eqref{Sdk-Closed}, which is equivalent to reduced interference as evident from~\eqref{Idk-ClosedNew}) while increasing the interference at the BS for UL signals (see~\eqref{Iuk-Closed}). This results in an increase in the DL SE and a decrease in the UL SE for both STARS- and cRIS-enabled systems, \textcolor{black}{which results in an increase of the total SE at the beginning, while a decrease is observed at high power.} Moreover,  the STARS-enabled system outperforms its cRIS-enabled counterpart in terms of sum SE \textcolor{black}{since the cRIS appears a decrease at high power.} It is interesting to note that increasing $p_{\mathrm b}$ does not necessarily increase the sum SE for both the systems due to equal power allocation for all the UEs at the BS. \textcolor{black}{This calls for the use of optimal power allocation at the BS, which is beyond the scope of this paper} and we consider the optimal transmit beamforming design for the system under consideration as a future work. A similar analysis is shown in Fig.~\ref{fig:ratePu} where we vary $p_{\mathrm u}$ with all other system parameters kept fixed. Analogous to the previous settings, increasing $p_{\mathrm u}$ increases the UL SINR $\gamma_{\mathrm uk}$ (see~\eqref{Suk-Closed} and~\eqref{Iuk-Closed}) while also increasing the DL interference (refer to~\eqref{Idk-ClosedNew}. This in turn increases the UL SE and decreases the DL SE for both STARS- and cRIS-aided systems, however, the STARS-assisted system outperforms its cRIS-assisted counterpart in terms of sum SE. Nonetheless, as shown in Fig.~\ref{fig:ratePu}, increasing $p_{\mathrm u}$ does not result in a monotonic increment in sum SE for the STARS-enabled system, which calls for the need for UL power control, and is beyond the scope of this paper. 
	
	In Fig.~\ref{fig:rateMr}, we depict the system performance for a varying number of BS receive antennas $M_{\mathrm R}$ while keeping all other system parameters fixed. It is clear from the figure that with a larger $M_{\mathrm R}$, the MRC at the BS results in a higher UL SE, whereas the DL SE remains unchanged. \textcolor{black}{It is interesting to note that the cRIS-enabled system achieves higher DL SE compared to that achieved by the STARS-enabled system because the inter-user interference in  STARS is higher since it includes all the users behind and in front of the surface. On the contrary, in the case of the cRIS, the users of the transmit part of the cRIS do not interfere with the users in reflective part of the cRIS, which means that the inter-user interference is lower and the rate is higher.} Analogously, in Fig.~\ref{fig:rateMt}, we show the dependence of the SE on varying the number of transmit antennas at the BS (i.e., $M_{\mathrm T}$). As expected, an increase in $M_{\mathrm T}$ results in a higher DL SE with UL SE being unchanged for both STARS- and cRIS-enabled systems. However, the STARS-enabled system outperforms the corresponding cRIS-assisted system in both  Figs.~\ref{fig:rateMr} and~\ref{fig:rateMt}. 
	
	\begin{figure}[t]
		\centering
		\includegraphics[width=0.7\linewidth]{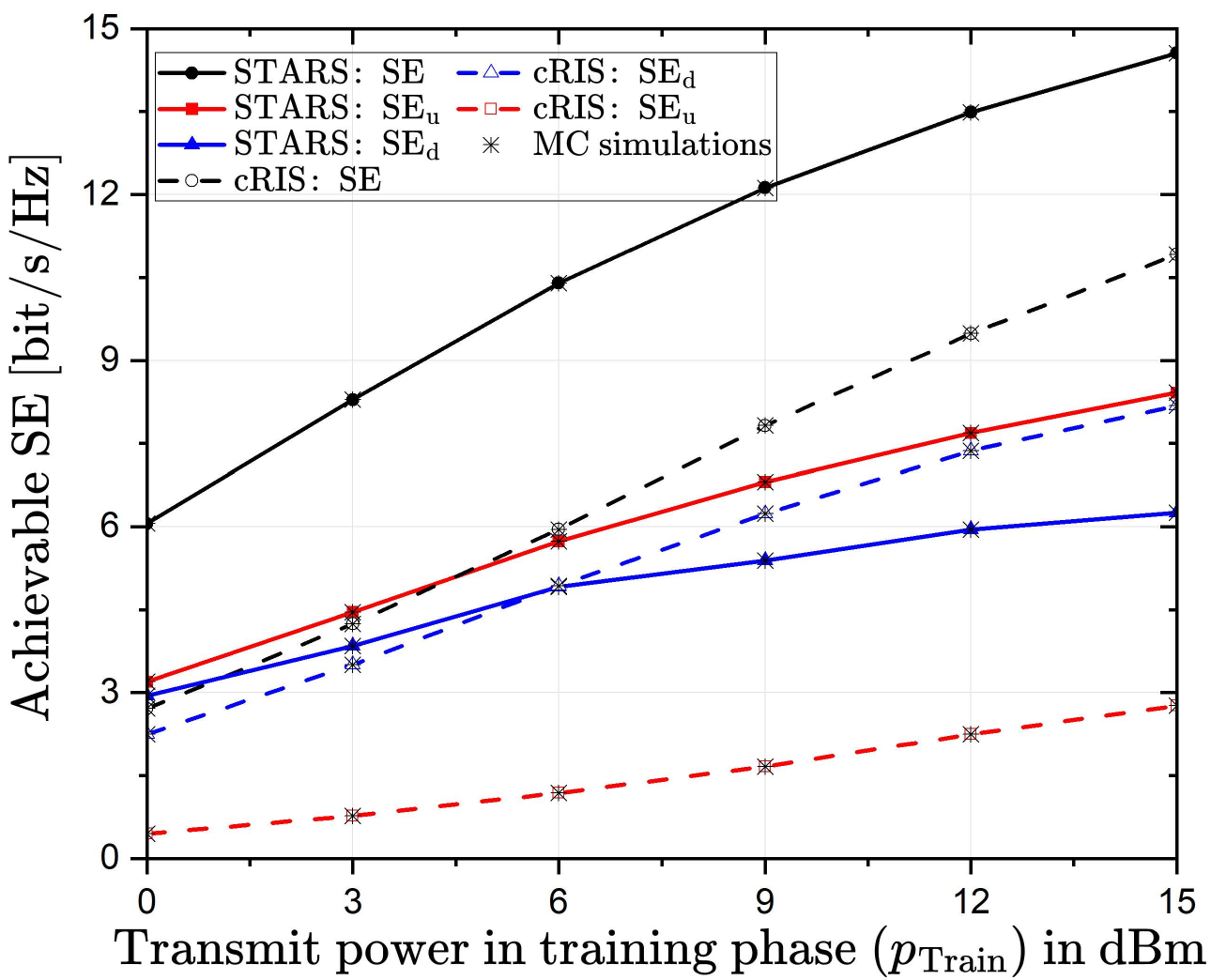}
		\caption{Impact of transmit power in the UL and DL training phase on the FD system performance.}
		\label{fig:ratePtrain}
	\end{figure}
	
	The impact of varying transmit power during the channel estimation phase, i.e., $p_{\mathrm{Train}}$ on the SE of both STARS- and cRIS-aided systems is illustrated in Fig~\ref{fig:ratePtrain}. Note that an increase of $p_{\mathrm{Train}}$ reduces the channel estimation error or enhances the quality of channel estimation, which in turn increases both UL and DL SE for both STARS- and cRIS-enabled systems as evident from the figure. However, the STARS-enabled system attains a significantly higher sum SE compared to the cRIS-assisted system. \textcolor{black}{Notably, in the downlink, initially, for low $p_{\mathrm{Train}}$, the channel estimation error 	is less significant in  STARS. However, as $p_{\mathrm{Train}}$ increases, the channel estimation error becomes more significant in STARS because this term is  included in the interference, which is higher in the downlink of STARS as explained before. This justifies the following behaviour, where the downlink of the cRIS system is initially inferior to the downlink of the STARS. However, as the pilot power increases beyond a certain threshold, the downlink of the cRIS shows superior performance and the gap widens with further increase in the pilot power.}
	
\begin{figure}[t]
		\centering
		\includegraphics[width=0.7\linewidth]{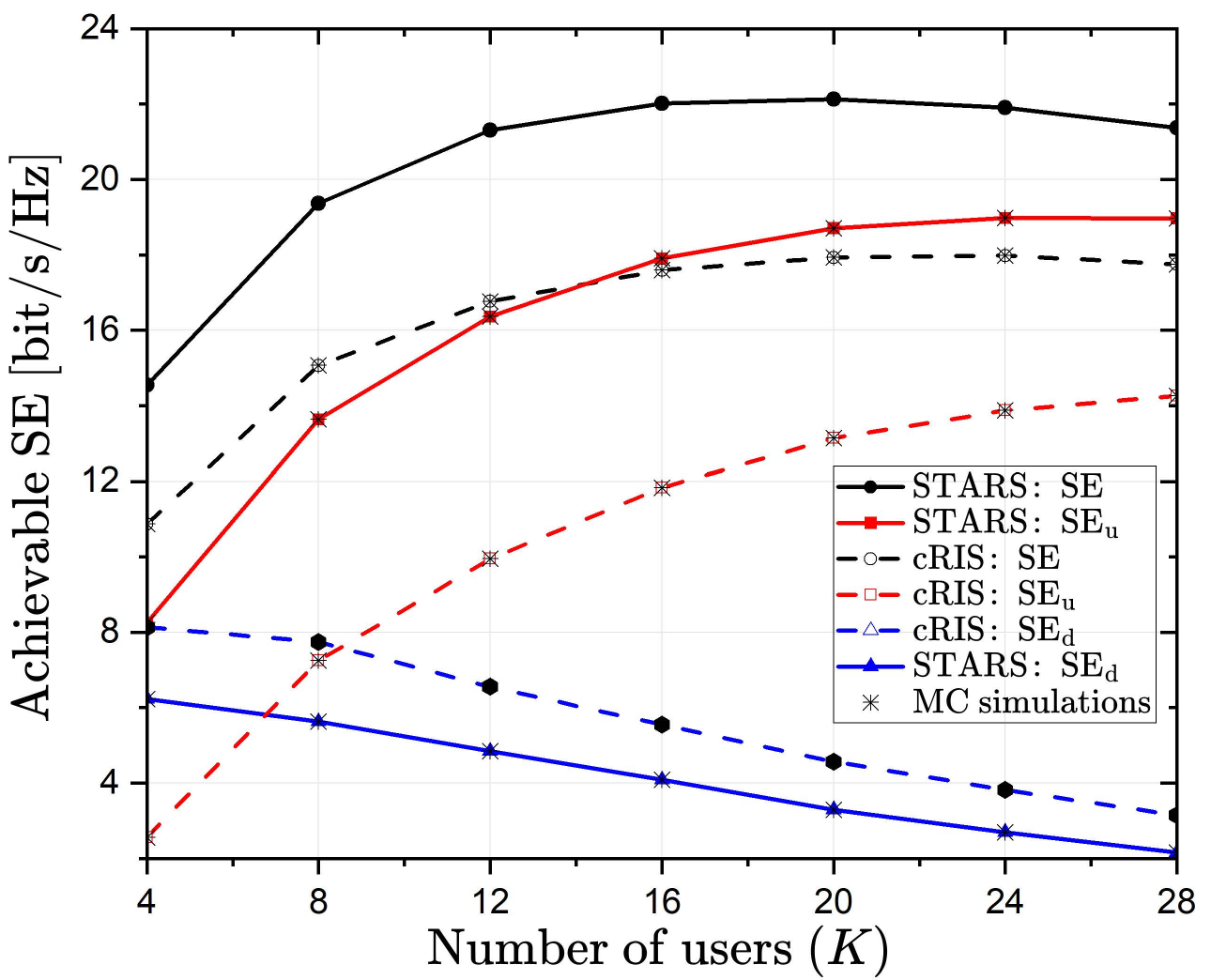}
		\caption{Impact of the number of users on the FD system performance.}
		\label{fig:rateK}
	\end{figure}

	In Fig.~\ref{fig:rateK}, we show the impact of increasing the number of UEs on the SE of STARS- and cRIS-enabled systems. Note that in this figure, we assume $K_{\mathrm r} = K_{\mathrm t} = K/2$. For both STARS- and cRIS-aided systems, it is clear from the figure that an increasing value of $K$ decreases the DL SE due to an increase in the DL interference. On the other hand, the UL SE first increases and then saturates for large $K$. However, more interestingly, the sum SE for both the systems first increases and then decreases, indicating that the system cannot serve a massive number of UEs with high SE without invoking optimal power control at the UEs.

	\section{Conclusion} \label{Conclusion} 
	This work has considered a novel multi-user STARS\textcolor{black}{-enabled mMIMO} deployment in an FD two-way communication network under the practical settings of correlated fading and imperfect CSI. We have modeled both the uplink and downlink channels of the considered system, \textcolor{black}{and also} derived the MMSE estimates of the effective channels. Next, we have derived closed-form expressions for \textcolor{black}{the average} uplink and downlink SE. \textcolor{black}{Furthermore, we have devised a numerically efficient projected gradient ascent method to obtain optimal STARS PBM structure to maximize the sum SE. The superiority of the STARS-enabled FD system over the corresponding cRIS-enabled FD system and STARS-enabled HD system was established via extensive numerical results.}
	
	%=================== APPENDICES==========================
	\appendices
	
	%--------------- Calculating UL-SINR
	\section{Proof of Theorem~\ref{thm:UL-SINR}} \label{sec:proof-UL-SINR}
	Using the fact that $\mathbf{x}\herm\mathbf{y}=\tr\big(\mathbf{y}\herm\mathbf{x}\big)$,~\eqref{Suk-Def} yields
	\begin{align}
		& S_{\mathrm{u}k} = p_{\mathrm{u}k}|\mathbb{E}\{\mathbf{v}_{k}\herm\tilde{\mathbf{u}}_{k}\}|^{2} = p_{\mathrm{u}k}|\tr\big(\mathbb{E}\big\{\tilde{\mathbf{u}}_{k}\mathbf{v}_{k}\herm\big\}\big)|^{2} \notag \\
		= & \ p_{\mathrm{u}k}|\tr\big(\mathbb{E}\big\{\tilde{\mathbf{u}}_{k}\hat{\tilde{\mathbf{u}}}_{k}\herm\big\}\big)|^{2} = p_{\mathrm{u}k}|\tr\big(\mathbb{E}\big\{\tilde{\mathbf{u}}_{k}\tilde{\mathbf{r}}_{k}\herm\tilde{\mathbf{Q}}_{k}\tilde{\mathbf{R}}_{k}\big\}\big)|^{2} \notag \\
		= & \ p_{\mathrm{u}k}|\tr\big(\mathbb{E}\big\{\tilde{\mathbf{u}}_{k}\tilde{\mathbf{u}}_{k}\herm\big\}\tilde{\mathbf{Q}}_{k}\tilde{\mathbf{R}}_{k}\big)|^{2} = p_{\mathrm{u}k}|\tr\big(\tilde{\boldsymbol{\Psi}}_{k}\big)|^{2} \notag \\
		= & \ p_{\mathrm{u}k}\tr^{2}\big(\tilde{\boldsymbol{\Psi}}_{k}\big). \label{Suk-Derivation}
	\end{align}
	Note that the last equality holds since $\tilde{\boldsymbol{\Psi}}_{k}$
	is Hermitian. The first term in~\eqref{Iuk-Def} leads to
	\begin{align}
		& \mathsf{var}\big\{\sqrt{p_{\mathrm{u}k}}\mathbf{v}_{k}\herm\tilde{\mathbf{u}}_{k}\big\} = p_{\mathrm{u}k}\mathbb{E}\big\{\big|\mathbf{v}_{k}\herm\tilde{\mathbf{u}}_{k}-\mathbb{E}\big\{\mathbf{v}_{k}\herm\tilde{\mathbf{u}}_{k}\big\}\big|^{2}\big\} \notag \\
		= & \ p_{\mathrm{u}k}\big[\mathbb{E}\big\{\big|\mathbf{v}_{k}\herm\tilde{\mathbf{u}}_{k}\big|^{2}\big\}-\big|\mathbb{E}\big\{\mathbf{v}_{k}\herm\tilde{\mathbf{u}}_{k}\big\}\big|^{2}\big] \notag \\
		= &\ p_{\mathrm{u}k}\big[\mathbb{E}\big\{\big|\hat{\tilde{\mathbf{u}}}_{k}\herm\tilde{\mathbf{u}}_{k}\big|^{2}\big\}-\big|\mathbb{E}\big\{\hat{\tilde{\mathbf{u}}}_{k}\herm\tilde{\mathbf{u}}_{k}\big\}\big|^{2}\big] \notag \\
		= & \ p_{\mathrm{u}k}\big[\mathbb{E}\big\{\big|\hat{\tilde{\mathbf{u}}}_{k}\herm\hat{\tilde{\mathbf{u}}}_{k}+\hat{\tilde{\mathbf{u}}}_{k}\herm\tilde{\mathbf{e}}_{k}\big|^{2}\big\}-\big|\mathbb{E}\big\{\hat{\tilde{\mathbf{u}}}_{k}\herm\tilde{\mathbf{u}}_{k}\big\}\big|^{2}\big] \notag \\
		= &\ p_{\mathrm{u}k}\mathbb{E}\big\{\big|\hat{\tilde{\mathbf{u}}}_{k}\herm\tilde{\mathbf{e}}_{k}\big|^{2}\big\}=p_{\mathrm{u}k}\mathbb{E}\big\{\tr\big(\hat{\tilde{\mathbf{u}}}_{k}\herm\tilde{\mathbf{e}}_{k}\tilde{\mathbf{e}}_{k}\herm\hat{\tilde{\mathbf{u}}}_{k}\big)\big\} \notag \\
		= & \ p_{\mathrm{u}k}\mathbb{E}\big\{\tr\big(\tilde{\mathbf{e}}_{k}\tilde{\mathbf{e}}_{k}\herm\hat{\tilde{\mathbf{u}}}_{k}\hat{\tilde{\mathbf{u}}}_{k}\herm\big)\big\} = p_{\mathrm{u}k}\tr\big(\big\{\tilde{\mathbf{R}}_{k}-\tilde{\boldsymbol{\Psi}}_{k}\big\}\tilde{\boldsymbol{\Psi}}_{k}\big) \notag \\
		= &\ p_{\mathrm{u}k}\big[\tr\big(\tilde{\boldsymbol{\Psi}}_{k}\tilde{\mathbf{R}}_{k}\big)-\tr\big(\tilde{\boldsymbol{\Psi}}_{k}^{2}\big)\big]. \label{Iuk1-Derivation}
	\end{align}
	Next, by using the independence between $\mathbf{v}_{k}\herm$ and $\tilde{\mathbf{u}}_{i}$, the  second term in~\eqref{Iuk-Def} yields
	\begin{align}
		& \sum_{i\in\mathcal{K}\setminus\{k\}}p_{\mathrm{u}i}\mathbb{E}\{|\mathbf{v}_{k}\herm\tilde{\mathbf{u}}_{i}|^{2}\}=\sum_{i\in\mathcal{K}\setminus\{k\}}p_{\mathrm{u}i}\mathbb{E}\{|\hat{\tilde{\mathbf{u}}}_{k}\herm\tilde{\mathbf{u}}_{i}|^{2}\} \notag \\
		= & \sum_{i\in\mathcal{K}\setminus\{k\}}p_{\mathrm{u}i}\mathbb{E}\{\tr(\hat{\tilde{\mathbf{u}}}_{k}\herm\tilde{\mathbf{u}}_{i}\tilde{\mathbf{u}}_{i}\herm\hat{\tilde{\mathbf{u}}}_{k})\} = \!\!\!\!\!\! \sum_{i\in\mathcal{K}\setminus\{k\}} \!\!\!\!\! p_{\mathrm{u}i}\mathbb{E}\{\tr(\hat{\tilde{\mathbf{u}}}_{k}\hat{\tilde{\mathbf{u}}}_{k}\herm\tilde{\mathbf{u}}_{i}\tilde{\mathbf{u}}_{i}\herm)\} \notag \\
		= & \sum_{i\in\mathcal{K}\setminus\{k\}}p_{\mathrm{u}i}\tr(\mathbb{E}\{\hat{\tilde{\mathbf{u}}}_{k}\hat{\tilde{\mathbf{u}}}_{k}\herm\}\mathbb{E}\{\tilde{\mathbf{u}}_{i}\tilde{\mathbf{u}}_{i}\herm\})=\!\!\!\!\sum_{i\in\mathcal{K}\setminus\{k\}}p_{\mathrm{u}i}\tr(\tilde{\boldsymbol{\Psi}}_{k}\tilde{\mathbf{R}}_{i}). \label{Iuk2-Derivation}
	\end{align}
	Similarly, the third term in~\eqref{Iuk-Def} is given by 
	\begin{align}
		& \dfrac{\beta p_{\mathrm{b}}}{K}\sum_{j\in\mathcal{K}}\mathbb{E}\{|\mathbf{v}_{k}\herm(\tilde{\mathbf{G}}\boldsymbol{\Theta}_{\mathrm{r}}\mathbf{G}+\mathbf{G}_{\mathrm{b}})\mathbf{f}_{j}|^{2}\} \notag \\
		= \ & \dfrac{\beta p_{\mathrm{b}}}{K}\sum_{j\in\mathcal{K}}\big[\mathbb{E}\{|\mathbf{v}_{k}\herm\tilde{\mathbf{G}}\boldsymbol{\Theta}_{\mathrm{r}}\mathbf{G}\mathbf{f}_{j}|^{2}\}+\mathbb{E}\{|\mathbf{v}_{k}\herm\mathbf{G}_{\mathrm{b}}\mathbf{f}_{j}|^{2}\}\big] \notag \\
		= \ &\ \dfrac{\beta p_{\mathrm{b}}}{K}\sum_{j\in\mathcal{K}}\big[\delta_{g}\tilde{\delta}_{g}\tr\big(\tilde{\boldsymbol{\Psi}}_{k}\tilde{\mathbf{R}}_{\mathrm{b}}\big)\tr\big(\b A_{\mathrm r}\boldsymbol{\Theta}_{\mathrm{r}}\herm\big)\tr\big(\mathbf{R}_{\mathrm{b}}\boldsymbol{\Psi}_{j}\big) \notag \\
		& +\sigma_{\mathrm{L}}^{2}\tr\big(\tilde{\boldsymbol{\Psi}}_{k}\tilde{\mathbf{R}}_{\mathrm{b}}\big)\tr\big(\boldsymbol{\Psi}_{j}\mathbf{R}_{\mathrm{b}}\big)\big]. \label{Iuk3-Derivation}
	\end{align}
	Moreover, the normalization term $\beta$ is given by 
	\begin{align}
		\beta = \ & \dfrac{K}{\sum_{\jmath \in\mathcal{K}}\mathbb{E}\big\{\mathbf{f}_{\jmath}\herm\mathbf{f}_{\jmath}\big\}} = \dfrac{K}{\sum_{\jmath \in\mathcal{K}}\mathbb{E}\big\{\hat{\mathbf{u}}_{\jmath}\hat{\mathbf{u}}_{\jmath}\herm\big\}} \notag \\
		= \ & \dfrac{K}{\sum_{\jmath \in\mathcal{K}}\tr(\boldsymbol{\Psi}_{\jmath})}=\dfrac{K}{\tr(\boldsymbol{\Psi}_{\mathrm{sum}})}. \label{betaClosed}
	\end{align}
	The last term in~\eqref{Iuk-Def} is given by 
	\begin{equation}
		\sigma_{\mathrm{b}}^{2}\mathbb{E}\big\{\big\Vert\mathbf{v}_{k}\herm\big\Vert^{2}\big\}=\sigma_{\mathrm{b}}^{2}\mathbb{E}\big\{\big\Vert\hat{\tilde{\mathbf{b}}}_{k}\herm\big\Vert^{2}\big\}=\sigma^2_{\mathrm b}\tr\big(\tilde{\boldsymbol{\Psi}}_{k}\big). \label{Iuk4-Derivation}
	\end{equation}
	Using~\eqref{Suk-Derivation} -- \eqref{Iuk4-Derivation}, closed-form expressions for $S_{\mathrm uk}$ and $I_{\mathrm uk}$ can directly be written as~\eqref{Suk-Closed} and~\eqref{Iuk-Closed}, respectively. This completes the proof.

	%--------------- Calculating DL-SINR
	\section{Proof of Theorem~\ref{thm:DL-SINR}} \label{sec:proof-DL-SINR}
	Since the BS uses MRT/conjugate beamforming based on the estimated CSI, we have $\mathbf{f}_{k}=\hat{\mathbf{u}}_{k}\herm$. Therefore,~\eqref{Sdk-Def} yields
	\begin{align}
		& S_{\mathrm{d}k} = \dfrac{p_{\mathrm{b}}}{\tr(\boldsymbol{\Psi}_{\mathrm{sum}})}|\mathbb{E}\{\mathbf{u}_{k}\mathbf{f}_{k}\}|^{2} = \dfrac{p_{\mathrm{b}}}{\tr(\boldsymbol{\Psi}_{\mathrm{sum}})}|\mathbb{E}\big\{\tr\big(\mathbf{f}_{k}\mathbf{u}_{k}\big)\big\}|^{2} \notag \\
		=  \ & \dfrac{p_{\mathrm{b}}}{\tr(\boldsymbol{\Psi}_{\mathrm{sum}})}|\mathbb{E}\big\{\tr\big(\hat{\mathbf{u}}_{k}\herm\mathbf{u}_{k}\big)\big\}|^{2} 
		= \dfrac{p_{\mathrm{b}}}{\tr(\boldsymbol{\Psi}_{\mathrm{sum}})}|\tr\big(\mathbb{E}\big\{\mathbf{R}_{k}\mathbf{Q}_{k}\mathbf{r}_{k}\herm\mathbf{u}_{k}\big\}\big)|^{2} \notag \\
		= \ &  \dfrac{p_{\mathrm{b}}}{\tr(\boldsymbol{\Psi}_{\mathrm{sum}})}|\tr\big(\mathbf{R}_{k}\mathbf{Q}_{k}\mathbb{E}\big\{\mathbf{u}_{k}\herm\mathbf{u}_{k}\big\}\big)|^{2} = \dfrac{p_{\mathrm{b}}}{\tr(\boldsymbol{\Psi}_{\mathrm{sum}})}\tr^{2}\big(\boldsymbol{\Psi}_{k}\big). \label{Sdk-Derivation}
	\end{align}
	Next, following the arguments in~Appendix~\ref{sec:proof-UL-SINR}, closed-form expressions for the first two terms in~\eqref{Idk-Def} can be expressed as $\frac{p_{\mathrm{b}}}{\tr(\boldsymbol{\Psi}_{\mathrm{sum}})}\big[\tr\big(\boldsymbol{\Psi}_{k}\mathbf{R}_{k}\big)-\tr\big(\boldsymbol{\Psi}_{k}^{2}\big)\big]$ and $\frac{p_{\mathrm{b}}}{\tr(\boldsymbol{\Psi}_{\mathrm{sum}})}\sum_{i\in\mathcal{K}\setminus\{k\}}\tr\big(\mathbf{R}_{k} \boldsymbol{\Psi}_{i}\big)$. The third term in~\eqref{Idk-Def} yields, 
	\begin{align}
		& \sum_{j\in\mathcal{K}}p_{\mathrm{u}j}\big(\sigma_{kj}^{2}+\mathbb{E}\big\{|\mathbf{h}_{k}\boldsymbol{\Theta}_{w_{k}}\tilde{\mathbf{h}}_{j}|^{2}\big\}) \notag \\
		= \ & \sum_{j\in\mathcal{K}}p_{\mathrm{u}j}\big[\sigma_{kj}^{2} + \mathbb{E}\big\{\tr\big(\mathbf{h}_{k}\boldsymbol{\Theta}_{w_{k}}\tilde{\mathbf{h}}_{j}\tilde{\mathbf{h}}_{j}\herm\boldsymbol{\Theta}_{w_{k}}\herm\mathbf{h}_{k}\herm\big)\big\}\big] \notag \\
		= \  & \sum_{j\in\mathcal{K}}p_{\mathrm{u}j}\big[\sigma_{kj}^{2} + \mathbb{E}\big\{\tr\big(\mathbf{h}_{k}\herm\mathbf{h}_{k}\boldsymbol{\Theta}_{w_{k}}\tilde{\mathbf{h}}_{j}\tilde{\mathbf{h}}_{j}\herm\boldsymbol{\Theta}_{w_{k}}\herm\big)\big\}\big] \notag \\
		= \ &  \sum_{j\in\mathcal{K}}p_{\mathrm{u}j}\big[\sigma_{kj}^{2} + \tr\big(\mathbb{E}\big\{\mathbf{h}_{k}\herm\mathbf{h}_{k}\big\}\boldsymbol{\Theta}_{w_{k}}\mathbb{E}\big\{\tilde{\mathbf{h}}_{j}\tilde{\mathbf{h}}_{j}\herm\big\}\boldsymbol{\Theta}_{w_{k}}\herm\big)\big] \notag \\
		=\  & \sum_{j\in\mathcal{K}}p_{\mathrm{u}j}\big[\sigma_{kj}^{2}+\delta_{h_{k}}\tilde{\delta}_{h_{j}}\tr\big(\mathbb{E}\big\{\mathbf{\mathbf{R}}_{\mathrm{s}}^{1/2}\mathbf{c}_{k}\herm\mathbf{c}_{k}\mathbf{R}_{\mathrm{s}}^{1/2}\big\} \notag \\
		& \qquad \qquad \times  \boldsymbol{\Theta}_{w_{k}} \mathbb{E}\big\{\mathbf{R}_{\mathrm{s}}^{1/2}\tilde{\mathbf{c}}_{j}\tilde{\mathbf{c}}_{j}\herm\mathbf{R}_{\mathrm{s}}^{1/2}\big\} \boldsymbol{\Theta}_{w_{k}}\herm\big)\big] \notag \\
		= \  & \sum_{j\in\mathcal{K}}p_{\mathrm{u}j}\big[\sigma_{kj}^{2} + \delta_{h_{k}}\tilde{\delta}_{h_{j}}\tr\big(\mathbf{R}_{\mathrm{s}}\boldsymbol{\Theta}_{w_{k}}\mathbf{R}_{\mathrm{s}}\boldsymbol{\Theta}_{w_{k}}\herm\big)\big] \notag \\
		= \ & \sum_{j\in\mathcal{K}}p_{\mathrm{u}j}\big[\sigma_{kj}^{2}+\delta_{h_{k}}\tilde{\delta}_{h_{j}}\tr\big(\b A_{w_k}\boldsymbol{\Theta}_{w_{k}}\herm\big)\big]. \label{Idk3-Derivation}
	\end{align}
	Therefore, using the fact that $z_{\mathrm dk} \sim \mathcal{CN}(0,\sigma^2_{\mathrm d})$,~\eqref{Sdk-Derivation},~\eqref{Idk3-Derivation}, and the arguments above, closed-form expressions for $S_{\mathrm dk}$ and $I_{\mathrm dk}$ are respectively given by~\eqref{Sdk-Closed} and~\eqref{Idk-Closed}. This completes the proof.

	%--------------- Calculating derivative of Suk wrt thetaR
	\section{Proof of Theorem~\ref{thm:gradClosed}} \label{sec:proof-gradClosed}
	Using~\eqref{Suk-Closed}, we have 
	\begin{align}
		& \nabla_{\boldsymbol{\theta}_{\mathrm m}} S_{\mathrm uk} = \ 2 p_{\mathrm uk} \tr \big( \tilde{\boldsymbol{\Psi}}_k \big) \tr \big\{ \nabla_{\boldsymbol{\theta}_{\mathrm m}} \big(\tilde{\mathbf R}_k \tilde{\mathbf Q}_k \tilde{\mathbf R}_k\big) \big\} \notag \\
		= & \ 2 p_{\mathrm uk} \tr \big( \tilde{\boldsymbol{\Psi}}_k \big) \tr \big\{ \nabla_{\boldsymbol{\theta}_{\mathrm m}} \big(\tilde{\mathbf R}_k \big) \tilde{\mathbf Q}_k \tilde{\mathbf R}_k \notag \\
		& \qquad \qquad -  \tilde{\mathbf R}_k \tilde{\mathbf Q}_k \nabla_{\boldsymbol{\theta}_{\mathrm m}} \big(\tilde{\mathbf R}_k \big) \tilde{\mathbf Q}_k \tilde{\mathbf R}_k + \tilde{\mathbf R}_k \tilde{\mathbf Q}_k \nabla_{\boldsymbol{\theta}_{\mathrm m}} \big(\tilde{\mathbf R}_k \big) \big\} \notag \\
		= & \  2 \tilde \delta_g \tilde \delta_{h_k} p_{\mathrm uk} \tr \big( \tilde{\boldsymbol{\Psi}}_k \big) \tr \big\{ \tilde{\mathbf R}_{\mathrm{b}} \tilde{\mathbf C}_k \nabla_{\boldsymbol \theta_{\mathrm m}}\big(\mathbf A_{\mathrm r} \boldsymbol \Theta_{\mathrm r}\herm \big)\big\} \notag \\
		= & \ \begin{cases} \tilde \nu_k \diag(\mathbf A_{\mathrm r}), & \text{if } \mathrm m = \mathrm r \\ \boldsymbol 0, & \text{otherwise } \end{cases}, \label{gradSuk-thetaM-Closed}
	\end{align}
	where $\tilde{\mathbf C}_k \triangleq \tilde{\mathbf Q}_k \tilde{\mathbf R}_k - \tilde{\mathbf Q}_k \tilde{\mathbf R}_k^2 \tilde{\mathbf Q}_k + \tilde{\mathbf R}_k \tilde{\mathbf Q}_k$, and $\tilde \nu_k \triangleq 2 \tilde \delta_g \tilde \gamma_{h_k} p_{\mathrm uk} \tr \big( \tilde{\boldsymbol{\Psi}}_k \big) \tr \big( \tilde{\mathbf R}_{\mathrm{b}} \tilde{\mathbf C}_k  \big)$. Analogously, closed-form expressions for $\nabla_{\boldsymbol{\theta}_{\mathrm r}} S_{\mathrm dk}$ and $\nabla_{\boldsymbol{\theta}_{\mathrm t}} S_{\mathrm dk}$ are given by~\eqref{gradSdk-thetaR-Closed}, and~\eqref{gradSdk-thetaT-Closed}, respectively. 
	
	Next, using~\eqref{Iuk-Closed}, let $I_{\mathrm{u}k}$ is given by~\eqref{Iuk-components}.
	\begin{figure*}
		\begin{align}
			I_{\mathrm{u}k} = & \ \underbrace{\tr\big(\tilde{\boldsymbol{\Psi}}_{k}\tilde{\mathbf{R}}_{\mathrm{sum}}\big)}_{I_{\mathrm{u}k1}} +\underbrace{\dfrac{p_{\mathrm{b}}}{\tr\big(\boldsymbol{\Psi}_{\mathrm{sum}}\big)}\tr\big(\tilde{\boldsymbol{\Psi}}_{k}\tilde{\mathbf{R}}_{\mathrm{b}}\big)\tr\big(\mathbf{R}_{\mathrm{b}}\boldsymbol{\Psi}_{\mathrm{sum}}\big)\big\{\delta_{g}\tilde{\delta}_{g}\tr\big(\b A_{\mathrm r} \boldsymbol{\Theta}_{\mathrm{r}}\herm\big)+\sigma_{\mathrm{L}}^{2}\big\}}_{I_{\mathrm{u}k2}} -\underbrace{p_{\mathrm{u}k}\tr\big(\tilde{\boldsymbol{\Psi}}_{k}^{2}\big)}_{I_{\mathrm{u}k3}}+\underbrace{\sigma^2_{\mathrm b}\tr\big(\tilde{\boldsymbol{\Psi}}_{k}\big)}_{I_{\mathrm{u}k4}}. \label{Iuk-components}
		\end{align}
		\hrulefill
	\end{figure*}
	Therefore, the gradient of the first term in~\eqref{Iuk-components} is given by 
	\begin{align}
		& \nabla_{\boldsymbol{\theta}_{\mathrm{m}}}I_{\mathrm{u}k1} \notag \\
		= \ & \tr\big\{\nabla_{\boldsymbol{\theta}_{\mathrm{m}}} \big(\tilde{\boldsymbol{\Psi}}_{k}\big)\tilde{\mathbf{R}}_{\mathrm{sum}}\big\} + \sum_{j\in\mathcal{K}} p_{\mathrm uj} \tr\big\{\tilde{\boldsymbol{\Psi}}_{k}\nabla_{\boldsymbol{\theta}_{\mathrm{m}}}\big(\tilde{\mathbf{R}}_{j}\big)\big\} \notag \\
		= & \tr\big\{\tilde{\mathbf{B}}_{k1}\nabla_{\boldsymbol{\theta}_{\mathrm{m}}}\big(\tilde{\mathbf{R}}_{k}\big)\big\} + \sum_{j\in\mathcal{K}}p_{\mathrm uj}\tr\big\{\tilde{\boldsymbol{\Psi}}_{k}\nabla_{\boldsymbol{\theta}_{\mathrm{m}}}\big(\tilde{\mathbf{R}}_{j}\big)\big\}, \label{grad-ThetaM-Iuk1}
	\end{align}
	where $\tilde{\mathbf{B}}_{k1} \triangleq \tilde{\mathbf{Q}}_{k}\tilde{\mathbf{R}}_{k}\tilde{\mathbf{R}}_{\mathrm{sum}}-\tilde{\mathbf{Q}}_{k}\tilde{\mathbf{R}}_{k}\tilde{\mathbf{R}}_{\mathrm{sum}}\tilde{\mathbf{R}}_{k}\tilde{\mathbf{Q}}_{k}+\tilde{\mathbf{R}}_{\mathrm{sum}}\tilde{\mathbf{R}}_{k}\tilde{\mathbf{Q}}_{k}$.
	The gradient of the second term in~\eqref{Iuk-components} is given by
	\begin{align}
		& \nabla_{\boldsymbol{\theta}_{\mathrm{m}}}I_{\mathrm{u}k2} \notag \\
		= \ & \nabla_{\boldsymbol{\theta}_{\mathrm{m}}}\frac{p_{\mathrm{b}}}{\tr\big(\boldsymbol{\Psi}_{\mathrm{sum}}\big)} \notag \\
		& \times \tr\big(\tilde{\boldsymbol{\Psi}}_{k}\tilde{\mathbf{R}}_{\mathrm{b}}\big)\tr\big(\mathbf{R}_{\mathrm{b}}\boldsymbol{\Psi}_{\mathrm{sum}}\big)\bigg\{\delta_{g}\tilde{\delta}_{g}\tr\big(\mathbf{A}_{\mathrm{r}}\boldsymbol{\Theta}_{\mathrm{r}}\herm\big)+\sigma_{\mathrm{L}}^{2}\bigg\} \notag \\
		= & \ \nabla_{\boldsymbol{\theta}_{\mathrm{m}}}\frac{p_{\mathrm{b}}\delta_{g}\tilde{\delta}_{g}}{\tr\big(\boldsymbol{\Psi}_{\mathrm{sum}}\big)}\tr\big(\tilde{\boldsymbol{\Psi}}_{k}\tilde{\mathbf{R}}_{\mathrm{b}}\big)\tr\big(\mathbf{R}_{\mathrm{b}}\boldsymbol{\Psi}_{\mathrm{sum}}\big)\tr\big(\mathbf{A}_{\mathrm{r}}\boldsymbol{\Theta}_{\mathrm{r}}\herm\big) \notag \\
		& +\nabla_{\boldsymbol{\theta}_{\mathrm{m}}}\frac{p_{\mathrm{b}}\sigma_{\mathrm{L}}^{2}}{\tr\big(\boldsymbol{\Psi}_{\mathrm{sum}}\big)}\tr\big(\tilde{\boldsymbol{\Psi}}_{k}\tilde{\mathbf{R}}_{\mathrm{b}}\big)\tr\big(\mathbf{R}_{\mathrm{b}}\boldsymbol{\Psi}_{\mathrm{sum}}\big) \notag \\
		= & \ \frac{p_{\mathrm{b}}\delta_{g}\tilde{\delta}_{g}}{\tr^{2}\big(\boldsymbol{\Psi}_{\mathrm{sum}}\big)}\bigg[ \!\! \tr\big(\boldsymbol{\Psi}_{\mathrm{sum}}\big)\nabla_{\boldsymbol{\theta}_{\mathrm{m}}}\bigg\{ \!\! \tr\big(\tilde{\boldsymbol{\Psi}}_{k}\tilde{\mathbf{R}}_{\mathrm{b}}\big)\tr\big(\mathbf{R}_{\mathrm{b}}\boldsymbol{\Psi}_{\mathrm{sum}}\big) \notag \\
		& \times \tr\big(\mathbf{A}_{\mathrm{r}}\boldsymbol{\Theta}_{\mathrm{r}}\herm\big)\bigg\} -\tr\big(\tilde{\boldsymbol{\Psi}}_{k}\tilde{\mathbf{R}}_{\mathrm{b}}\big)\tr\big(\mathbf{R}_{\mathrm{b}}\boldsymbol{\Psi}_{\mathrm{sum}}\big) \tr\big(\mathbf{A}_{\mathrm{r}}\boldsymbol{\Theta}_{\mathrm{r}}\herm\big) \notag \\
		& \times \nabla_{\boldsymbol{\theta}_{\mathrm{m}}}\tr\big(\boldsymbol{\Psi}_{\mathrm{sum}}\big)\bigg]  + \frac{p_{\mathrm{b}}\sigma_{\mathrm{L}}^{2}}{\tr^{2}\big(\boldsymbol{\Psi}_{\mathrm{sum}}\big)}\bigg[\tr\big(\boldsymbol{\Psi}_{\mathrm{sum}}\big) \notag \\
		& \times \nabla_{\boldsymbol{\theta}_{\mathrm{m}}}\bigg\{\tr\big(\tilde{\boldsymbol{\Psi}}_{k}\tilde{\mathbf{R}}_{\mathrm{b}}\big)\tr\big(\mathbf{R}_{\mathrm{b}}\boldsymbol{\Psi}_{\mathrm{sum}}\big)\bigg\} -\tr\big(\tilde{\boldsymbol{\Psi}}_{k}\tilde{\mathbf{R}}_{\mathrm{b}}\big) \notag \\
		&  \times \tr\big(\mathbf{R}_{\mathrm{b}}\boldsymbol{\Psi}_{\mathrm{sum}}\big)\nabla_{\boldsymbol{\theta}_{\mathrm{m}}}\tr\big(\boldsymbol{\Psi}_{\mathrm{sum}}\big)\bigg] \notag \\
		= & \ \frac{p_{\mathrm{b}}\delta_{g}\tilde{\delta}_{g}}{\tr^{2}\big(\boldsymbol{\Psi}_{\mathrm{sum}}\big)}{\bigg[}\tr\big(\boldsymbol{\Psi}_{\mathrm{sum}}\big)\tr\big(\mathbf{R}_{\mathrm{b}}\boldsymbol{\Psi}_{\mathrm{sum}}\big)\tr\big(\mathbf{A}_{\mathrm{r}}\boldsymbol{\Theta}_{\mathrm{r}}\herm\big) \notag \\
		& \times \tr\big\{\big({\nabla_{\boldsymbol{\theta}_{\mathrm{m}}}\tilde{\boldsymbol{\Psi}}_{k}}\big)\tilde{\mathbf{R}}_{\mathrm{b}}\big\} + \tr\big(\boldsymbol{\Psi}_{\mathrm{sum}}\big)\tr\big(\tilde{\boldsymbol{\Psi}}_{k}\tilde{\mathbf{R}}_{\mathrm{b}}\big)  \notag \\
		& \times  \tr\big(\mathbf{A}_{\mathrm{r}}\boldsymbol{\Theta}_{\mathrm{r}}\herm\big) \tr\big\{\mathbf{R}_{\mathrm{b}}\big({\sum_{j\in\mathcal{K}}\nabla_{\boldsymbol{\theta}_{\mathrm{m}}}\boldsymbol{\Psi}_{j}}\big)\big\}+\tr\big(\boldsymbol{\Psi}_{\mathrm{sum}}\big) \notag \\
		& \times \! \tr \! \big(\tilde{\boldsymbol{\Psi}}_{k}\tilde{\mathbf{R}}_{\mathrm{b}}\big) \! \tr \! \big(\mathbf{R}_{\mathrm{b}}\boldsymbol{\Psi}_{\mathrm{sum}}\big) \!\nabla_{\boldsymbol{\theta}_{\mathrm{m}}} \! \tr\big(\mathbf{A}_{\mathrm{r}}\boldsymbol{\Theta}_{\mathrm{r}}\herm\big) \!-\!\tr\big(\tilde{\boldsymbol{\Psi}}_{k}\tilde{\mathbf{R}}_{\mathrm{b}}\big)  \notag \\
		&  \times \tr\big(\mathbf{R}_{\mathrm{b}}\boldsymbol{\Psi}_{\mathrm{sum}}\big) \tr\big(\mathbf{A}_{\mathrm{r}}\boldsymbol{\Theta}_{\mathrm{r}}\herm\big)\nabla_{\boldsymbol{\theta}_{\mathrm{m}}}\tr\big\{{\sum_{j\in\mathcal{K}}\nabla_{\boldsymbol{\theta}_{\mathrm{m}}}\boldsymbol{\Psi}_{j}}\big\}{\bigg]} \notag \\
		& +\frac{p_{\mathrm{b}}\sigma_{\mathrm{L}}^{2}}{\tr^{2}\big(\boldsymbol{\Psi}_{\mathrm{sum}}\big)}{\bigg[}\!\!\tr \! \big(\boldsymbol{\Psi}_{\mathrm{sum}}\big) \! \tr\big(\mathbf{R}_{\mathrm{b}}\boldsymbol{\Psi}_{\mathrm{sum}}\big) \tr\big\{\big({\nabla_{\boldsymbol{\theta}_{\mathrm{m}}}\tilde{\boldsymbol{\Psi}}_{k}}\big)\tilde{\mathbf{R}}_{\mathrm{b}}\big\} \notag \\
		&  + \! \tr \!\big(\boldsymbol{\Psi}_{\mathrm{sum}}\big) \! \tr \!\big(\tilde{\boldsymbol{\Psi}}_{k}\tilde{\mathbf{R}}_{\mathrm{b}}\big)\!\tr\!\big\{\mathbf{R}_{\mathrm{b}}\big({\sum_{j\in\mathcal{K}}\nabla_{\boldsymbol{\theta}_{\mathrm{m}}}\boldsymbol{\Psi}_{j}}\big)\big\} \notag \\
		&  -\tr\big(\tilde{\boldsymbol{\Psi}}_{k}\tilde{\mathbf{R}}_{\mathrm{b}}\big)\tr\big(\mathbf{R}_{\mathrm{b}}\boldsymbol{\Psi}_{\mathrm{sum}}\big)\nabla_{\boldsymbol{\theta}_{\mathrm{m}}}\tr\big\{{\sum_{j\in\mathcal{K}}\nabla_{\boldsymbol{\theta}_{\mathrm{m}}}\boldsymbol{\Psi}_{j}}\big\}{\bigg]} \notag \\
		= & \ \tr\big\{\tilde{\mathbf{B}}_{k2}\big(\nabla_{\boldsymbol{\theta}_{\mathrm{m}}}\tilde{\mathbf{R}}_{k}\big)\big\}+\sum_{j\in\mathcal{K}}\tr\big\{\tilde{\mathbf{L}}_{kj}\big(\nabla_{\boldsymbol{\theta}_{\mathrm{m}}}\mathbf{R}_{j}\big)\big\} \notag \\
		& +\tilde{\chi}_{k}\nabla_{\boldsymbol{\theta}_{\mathrm{m}}}\tr\big(\mathbf{A}_{\mathrm{r}}\boldsymbol{\Theta}_{\mathrm{r}}\herm\big), \label{grad-ThetaM-Iuk2}
	\end{align}
	where $\varpi \triangleq \frac{p_{\mathrm{b}}}{\tr\big(\boldsymbol{\Psi}_{\mathrm{sum}}\big)}\tr\big(\mathbf{R}_{\mathrm{b}}\boldsymbol{\Psi}_{\mathrm{sum}}\big)\big\{\delta_{g}\tilde{\delta}_{g}\tr\big(\mathbf{A}_{\mathrm{r}}\boldsymbol{\Theta}_{\mathrm{r}}\herm\big)+\sigma_{\mathrm{L}}^{2}\big\}$, $\tilde{\boldsymbol{\Xi}}_{k} \triangleq \frac{p_{\mathrm{b}}}{\tr\big(\boldsymbol{\Psi}_{\mathrm{sum}}\big)}\tr\big(\tilde{\boldsymbol{\Psi}}_{k}\tilde{\mathbf{R}}_{\mathrm{b}}\big)\big\{\delta_{g}\tilde{\delta}_{g}\tr\big(\mathbf{A}_{\mathrm{r}}\boldsymbol{\Theta}_{\mathrm{r}}\herm\big)+\sigma_{\mathrm{L}}^{2}\big\}\big[\mathbf{R}_{\mathrm{b}}-\frac{\tr\big(\mathbf{R}_{\mathrm{b}}\boldsymbol{\Psi}_{\mathrm{sum}}\big)}{\tr\big(\boldsymbol{\Psi}_{\mathrm{sum}}\big)}\big]$, $\tilde{\chi}_{k}\triangleq\frac{p_{\mathrm{b}}\delta_{g}\tilde{\delta}_{g}}{\tr\big(\boldsymbol{\Psi}_{\mathrm{sum}}\big)}\tr\big(\tilde{\boldsymbol{\Psi}}_{k}\tilde{\mathbf{R}}_{\mathrm{b}}\big)\tr\big(\mathbf{R}_{\mathrm{b}}\boldsymbol{\Psi}_{\mathrm{sum}}\big)$, $\tilde{\mathbf{B}}_{k2} \triangleq \varpi \big(\tilde{\mathbf{Q}}_{k}\tilde{\mathbf{R}}_{k}\tilde{\mathbf{R}}_{\mathrm{b}}-\tilde{\mathbf{Q}}_{k}\tilde{\mathbf{R}}_{k}\tilde{\mathbf{R}}_{\mathrm{b}}\tilde{\mathbf{R}}_{k}\tilde{\mathbf{Q}}_{k}+\tilde{\mathbf{R}}_{\mathrm{b}}\tilde{\mathbf{R}}_{k}\tilde{\mathbf{Q}}_{k}\big)$, and $\tilde{\mathbf{L}}_{kj} \triangleq \mathbf{Q}_{j}\mathbf{R}_{j}\tilde{\boldsymbol{\Xi}}_{k}-\mathbf{Q}_{j}\mathbf{R}_{j}\tilde{\boldsymbol{\Xi}}_{k}\mathbf{R}_{j}\mathbf{Q}_{j}+\tilde{\boldsymbol{\Xi}}_{k}\mathbf{R}_{j}\mathbf{Q}_{j}$.
	Similarly, the gradient of the third term in~\eqref{Iuk-components} can be obtained as 
	\begin{align}
		\nabla_{\boldsymbol{\theta}_{\mathrm{m}}}I_{\mathrm{u}k3} = \ & p_{\mathrm{u}k}\nabla_{\boldsymbol{\theta}_{\mathrm{m}}}\tr\big(\tilde{\boldsymbol{\Psi}}_{k}^{2}\big) = p_{\mathrm{u}k}\tr\big\{2\tilde{\boldsymbol{\Psi}}_{k}\big(\nabla_{\boldsymbol{\theta}_{\mathrm{m}}}\tilde{\boldsymbol{\Psi}}_{k}\big)\big\} \notag \\
		= \ & \tr\big\{\tilde{\mathbf{B}}_{k3}\nabla_{\boldsymbol{\theta}_{m}}\big(\tilde{\mathbf{R}}_{k}\big)\big\}, \label{grad-ThetaM-Iuk3}
	\end{align}
	where $\tilde{\mathbf{B}}_{k3}=2p_{\mathrm{u}k}\big(\tilde{\mathbf{Q}}_{k}\tilde{\mathbf{R}}_{k}\tilde{\boldsymbol{\Psi}}_{k}-\tilde{\mathbf{Q}}_{k}\tilde{\mathbf{R}}_{k}\tilde{\boldsymbol{\Psi}}_{k}\tilde{\mathbf{R}}_{k}\tilde{\mathbf{Q}}_{k}+\tilde{\boldsymbol{\Psi}}_{k}\tilde{\mathbf{R}}_{k}\tilde{\mathbf{Q}}_{k}\big)$. Moreover, the derivative of the last term in~\eqref{Iuk-components} can be computed as 
	\begin{align}
		\nabla_{\boldsymbol{\theta}_{\mathrm{m}}}I_{\mathrm{u}k4} & =\sigma^2_{\mathrm b}\nabla_{\boldsymbol{\theta}_{\mathrm{m}}}\tr\big(\tilde{\boldsymbol{\Psi}}_{k}\big)=\sigma^2_{\mathrm b}\tr\big\{\tilde{\mathbf{C}}_{k}\nabla_{\boldsymbol{\theta}_{m}}\big(\tilde{\mathbf{R}}_{k}\big)\big\}. \label{grad-ThetaM-Iuk4}
	\end{align}
	Using~\eqref{grad-ThetaM-Iuk1} -- \eqref{grad-ThetaM-Iuk4} with some algebraic manipulations yields
	\begin{align}
		& \nabla_{\boldsymbol{\theta}_{\mathrm{m}}}I_{\mathrm{u}k} \notag \\
		= \ &   \tilde{\delta}_{g}\tilde{\delta}_{h_{k}}\tr\big\{{\tilde{\mathbf{R}}_{\mathrm{b}}\big(\tilde{\mathbf{B}}_{k1}+\tilde{\mathbf{B}}_{k2}-\tilde{\mathbf{B}}_{k3}+\sigma_{\mathrm{u}}^{2}\tilde{\mathbf{C}}_{k}\big)}\nabla_{\boldsymbol{\theta}_{\mathrm{m}}}\big(\mathbf{A}_{\mathrm{r}}\boldsymbol{\Theta}_{\mathrm{r}}\herm\big)\big\} \notag \\
		& +\tilde{\delta}_{g}\sum_{j\in\mathcal{K}}\tilde{\delta}_{h_{j}} p_{\mathrm uj} \tr\big\{{\tilde{\mathbf{R}}_{\mathrm{b}}\tilde{\boldsymbol{\Psi}}_{k}}\nabla_{\boldsymbol{\theta}_{\mathrm{m}}}\big(\mathbf{A}_{\mathrm{r}}\boldsymbol{\Theta}_{\mathrm{r}}\herm\big)\big\} \notag \\ 
		& +\delta_{g}\sum_{j\in\mathcal{K}}\delta_{h_{j}}\tr\big\{{\mathbf{R}_{\mathrm{b}}\tilde{\mathbf{L}}_{kj}}\nabla_{\boldsymbol{\theta}_{\mathrm{m}}}\big(\mathbf{A}_{w_{j}}\boldsymbol{\Theta}_{w_{j}}\herm\big)\big\} \notag \\
		&  +\tilde{\chi}_{k}\nabla_{\boldsymbol{\theta}_{\mathrm{m}}}\tr\big(\mathbf{A}_{\mathrm{r}}\boldsymbol{\Theta}_{\mathrm{r}}\herm\big). \label{grad-ThetaM-Iuk}
	\end{align}
	Using~\eqref{grad-ThetaM-Iuk}, the closed-form expressions for $\nabla_{\boldsymbol{\theta}_{\mathrm{r}}}I_{\mathrm{u}k}$ and $\nabla_{\boldsymbol{\theta}_{\mathrm{t}}}I_{\mathrm{u}k}$ are respectively given by~\eqref{grad-thetaR-Iuk-Closed} and~\eqref{grad-thetaT-Iuk-Closed}.
	
	Turning our attention to the DL signaling, from~\eqref{Idk-ClosedNew} we write $I_{\mathrm{d}k}$ as given in~\eqref{Idk-components}.
	\begin{figure*}
		\begin{align}
			I_{\mathrm{d}k} & =\underbrace{\tr\big(\mathbf{R}_{k}\boldsymbol{\Psi}_{\mathrm{sum}}\big)}_{I_{\mathrm{d}k1}}+\underbrace{\frac{\tr\big(\boldsymbol{\Psi}_{\mathrm{sum}}\big)}{p_{\mathrm{b}}}\sum_{j\in\mathcal{K}}\Big[p_{\mathrm{u}k}\big\{\sigma_{kj}^{2}+\delta_{h_{k}}\tilde{\delta}_{h_{j}}\tr\big(\b A_{w_k} \boldsymbol{\Theta}_{w_{k}}\herm\big)\big\}\Big]}_{I_{\mathrm{d}k2}} -\underbrace{\tr\big(\boldsymbol{\Psi}_{k}^{2}\big)}_{I_{\mathrm{d}k3}}+\underbrace{\frac{\sigma_{\mathrm{d}}^{2}}{p_{\mathrm{b}}}\tr\big(\boldsymbol{\Psi}_{\mathrm{sum}}\big)}_{I_{\mathrm{d}k4}}. \label{Idk-components}
		\end{align}
		\hrulefill 
	\end{figure*}
	Therefore, the gradient of the first term in~\eqref{Idk-components} is given by
	\begin{align}
		& \nabla_{\boldsymbol{\theta}_{\mathrm{m}}}I_{\mathrm{d}k1} =  \nabla_{\boldsymbol{\theta}_{\mathrm{m}}}\tr\big(\boldsymbol{\Psi}_{\mathrm{sum}}\mathbf{R}_{k}\big) \notag \\
		= & \ \sum_{j\in\mathcal{K}}\Big[\tr\big\{\nabla_{\boldsymbol{\theta}_{\mathrm{m}}}\big(\boldsymbol{\Psi}_{j}\big)\mathbf{R}_{k}\big\}+\tr\big(\boldsymbol{\Psi}_{j}\nabla_{\boldsymbol{\theta}_{\mathrm{m}}}\big(\mathbf{R}_{k}\big)\big\}\Big] \notag \\
		= & \ \sum_{j\in\mathcal{K}}\tr\big\{\mathbf{L}_{kj}\nabla_{\boldsymbol{\theta}_{m}}\big(\mathbf{R}_{j}\big)\big\}+\tr\big\{\boldsymbol{\Psi}_{\mathrm{sum}}\nabla_{\boldsymbol{\theta}_{\mathrm{m}}}\big(\mathbf{R}_{k}\big)\big\}, \label{grad-ThetaM-Idk1}
	\end{align}
	where $\mathbf{L}_{kj}=\mathbf{Q}_{j}\mathbf{R}_{j}\mathbf{R}_{k}-\mathbf{Q}_{j}\mathbf{R}_{j}\mathbf{R}_{k}\mathbf{R}_{j}\mathbf{Q}_{j}+\mathbf{R}_{k}\mathbf{R}_{j}\mathbf{Q}_{j}$.
	The gradient of the second term in~\eqref{Idk-components} can be obtained as 
	\begin{align}
		& \nabla_{\boldsymbol{\theta}_{\mathrm{m}}}I_{\mathrm{d}k2} \notag \\ 
		= \ & \frac{1}{p_{\mathrm{b}}}\sum_{j\in\mathcal{K}}p_{\mathrm{u}j}\sigma_{kj}^{2}\tr\big(\nabla_{\boldsymbol{\theta}_{\mathrm{m}}}\boldsymbol{\Psi}_{\mathrm{sum}}\big) \notag \\
		& + \frac{\delta_{h_{k}}}{p_{\mathrm{b}}}\sum_{j\in\mathcal{K}}p_{\mathrm{u}j}\tilde{\delta}_{h_{j}}\nabla_{\boldsymbol{\theta}_{\mathrm{m}}}\tr\big(\boldsymbol{\Psi}_{\mathrm{sum}}\big) \tr\big(\b A_{w_k}\boldsymbol{\Theta}_{w_{k}}\herm\big) \notag \\
		= \ &  \frac{1}{p_{\mathrm{b}}}\sum_{j\in\mathcal{K}}p_{\mathrm{u}j}\Big[\sigma_{kj}^{2}+\delta_{h_{k}}\tilde{\delta}_{h_{j}}\tr\big(\mathbf{A}_{w_{k}}\boldsymbol{\Theta}_{w_{k}}\herm\big)\Big] \notag \\
		& \times \Big\{\sum\nolimits _{\imath\in\mathcal{K}}\tr\big\{\mathbf{C}_{\imath}\nabla_{\boldsymbol{\theta}_{\mathrm{m}}}\big(\mathbf{R}_{\imath}\big)\big\}\Big\} \notag \\ 
		& +\frac{\delta_{h_{k}}}{p_{\mathrm{b}}}\tr\big(\boldsymbol{\Psi}_{\mathrm{sum}}\big)\sum_{j\in\mathcal{K}}p_{\mathrm{u}j}\tilde{\delta}_{h_{j}}\nabla_{\boldsymbol{\theta}_{\mathrm{m}}}\tr\big(\mathbf{A}_{w_{k}}\boldsymbol{\Theta}_{w_{k}}\herm\big). \label{grad-ThetaM-Idk2}
	\end{align}
	Next, the gradient of the third term in~\eqref{Idk-components} is given by 
	\begin{equation}
		\nabla_{\boldsymbol{\theta}_{\mathrm{m}}}I_{\mathrm{d}k3}=\nabla_{\boldsymbol{\theta}_{\mathrm{m}}}\tr\big(\boldsymbol{\Psi}_{k}^{2}\big) = \tr\{\mathbf{B}_{k}\nabla_{\boldsymbol{\theta}_{\mathrm{m}}}\big(\mathbf{R}_{k}\big)\big\}, \label{grad-ThetaM-Idk3}
	\end{equation}
	where $\mathbf{B}_{k}=2\big(\mathbf{Q}_{k}\mathbf{R}_{k}\boldsymbol{\Psi}_{k}-\mathbf{Q}_{k}\mathbf{R}_{k}\boldsymbol{\Psi}_{k}\mathbf{R}_{k}\mathbf{Q}_{k}+\boldsymbol{\Psi}_{k}\mathbf{R}_{k}\mathbf{Q}_{k}\big)$.
	Additionally, the gradient of the last term in~\eqref{Idk-components} is obtained as 
	\begin{align}
		\nabla_{\boldsymbol{\theta}_{\mathrm{m}}}I_{\mathrm{d}k4} = \ & \frac{\sigma_{\mathrm{d}}^{2}}{p_{\mathrm{b}}}\tr\big(\sum\nolimits_{\imath\in\mathcal{K}}\nabla_{\boldsymbol{\theta}_{\mathrm{m}}}\boldsymbol{\Psi}_{\imath}\big) \notag \\
		= \ & \frac{\sigma_{\mathrm{d}}^{2}}{p_{\mathrm{b}}}\sum\nolimits _{\imath\in\mathcal{K}}\tr\big\{\mathbf{C}_{\imath}\nabla_{\boldsymbol{\theta}_{\mathrm{m}}}\big(\mathbf{R}_{\imath}\big)\big\}. \label{grad-ThetaM-Idk4}
	\end{align}
	Using some algebraic manipulations on~\eqref{Idk-components} -- \eqref{grad-ThetaM-Idk4}, the gradient of $I_{\mathrm dk}$ is given by 
	\begin{align}
		& \nabla_{\boldsymbol{\theta}_{\mathrm{m}}}I_{\mathrm{d}k} =\tr\Big\{\big(\boldsymbol{\Psi}_{\mathrm{sum}}-\mathbf{B}_{k}\big)\nabla_{\boldsymbol{\theta}_{\mathrm{m}}}\big(\mathbf{R}_{k}\big)\Big\} \notag \\
		& +\sum_{j\in\mathcal{K}}\tr\big\{\big(\mathbf{L}_{kj}\!+\!\chi_{k1}\b C_j \big) \nabla_{\boldsymbol{\theta}_{\mathrm{m}}}\big(\mathbf{R}_{j}\big)\big\}\!+\!\chi_{k2}\nabla_{\boldsymbol{\theta}_{\mathrm{m}}} \! \tr\big(\mathbf{A}_{w_{k}}\boldsymbol{\Theta}_{w_{k}}\herm\big) \notag \\
		& =\delta_{g}\delta_{h_{k}}\tr\big\{\mathbf{R}_{\mathrm{b}}\big(\boldsymbol{\Psi}_{\mathrm{sum}}-\mathbf{B}_{k}\big)\nabla_{\boldsymbol{\theta}_{\mathrm{m}}}\big(\mathbf{A}_{w_{k}}\boldsymbol{\Theta}_{w_{k}}\herm\big)\big\}\notag \\
		&  +\delta_{g}\sum_{j\in\mathcal{K}}\delta_{h_{j}}\tr\big\{\mathbf{R}_{\mathrm{b}}\big(\mathbf{L}_{kj}+\chi_{k1}\b C_j \big)\nabla_{\boldsymbol{\theta}_{\mathrm{m}}}\big(\mathbf{A}_{w_{j}}\boldsymbol{\Theta}_{w_{j}}\herm\big)\big\} \notag \\
		& +\chi_{k2}\nabla_{\boldsymbol{\theta}_{\mathrm{m}}}\tr\big(\mathbf{A}_{w_{k}}\boldsymbol{\Theta}_{w_{k}}\herm\big), \label{grad-ThetaM-Idk}
	\end{align}
	where $\boldsymbol{\Xi}_{k}=\boldsymbol{\Psi}_{\mathrm{sum}}-2\mathbf{T}_{k}$,
	$\chi_{k1}=\frac{1}{p_{\mathrm{b}}}\sum_{j\in\mathcal{K}}p_{\mathrm{u}j}\Big\{\sigma_{kj}^{2}+\delta_{h_{k}}\tilde{\delta}_{h_{j}}\tr\big(\mathbf{A}_{w_{k}}\boldsymbol{\Theta}_{w_{k}}\herm\big)\Big\}+\frac{\sigma_{\mathrm{d}}^{2}}{p_{\mathrm{b}}}$,
	and $\chi_{k2}=\frac{\delta_{h_{k}}}{p_{\mathrm{b}}}\tr\big(\boldsymbol{\Psi}_{\mathrm{sum}}\big)\sum_{j\in\mathcal{K}}p_{\mathrm{u}j}\tilde{\delta}_{h_{j}}$. 
	From~\eqref{grad-ThetaM-Idk}, closed-form expressions for $\nabla_{\b \theta_{\mathrm r}}I_{\mathrm dk}$ and $\nabla_{\b \theta_{\mathrm t}}I_{\mathrm dk}$ are respectively given by~\eqref{grad-thetaR-Idk-Closed} and~\eqref{grad-thetaT-Idk-Closed}. This completes the proof.

	%==============================
	\bibliographystyle{IEEEtran}
	\bibliography{IEEEabrv,mybib}
\end{document}